\documentclass[structabstract]{aa}  
                     
\usepackage{txfonts}
\usepackage{graphicx}
\usepackage[authoryear]{natbib}
\usepackage{color}

\usepackage{txfonts}
\newcommand{\hel}[2] {He\,{\sc #1}~$\rm{\lambda}$#2}

\newcommand{\hhel}[1] {He\,{\sc #1}}

\newcommand{\msun}{\ensuremath{\mathit{M}_{\odot}}}

\newcommand{\vrot}{\ensuremath{\varv_{\rm e}\sin i}}
\newcommand{\veq}{\ensuremath{\varv_{\rm e}}}

\def\kms{\mbox{${\rm km}\:{\rm s}^{-1}$}}

\def\lesssim{\mathrel{\hbox{\rlap{\hbox{\lower4pt\hbox{$\sim$}}}\hbox{$<$}}}}

\def\gtrsim{\mathrel{\hbox{\rlap{\hbox{\lower4pt\hbox{$\sim$}}}\hbox{$>$}}}}
\let\ga=\gtrsim

\defcitealias{evans}{Paper~I}
\defcitealias{dufton}{Paper~X}
\defcitealias{sana}{Paper~VIII}
\defcitealias{ramirezagudelo}{Paper~XII}
\defcitealias{doran}{Paper~XI}
\defcitealias{walborn2014}{Paper~XIV}

\titlerunning{VFTS~XXI: Stellar spin rates of O-type spectroscopic binaries }
\begin{document}
  \title{The VLT-FLAMES Tarantula Survey\thanks{Based on observations
   collected at the European Southern Observatory under program ID 182.D-0222.} 
}

   \subtitle{XXI. {Stellar spin rates of  O-type spectroscopic binaries}}
   \author{
     O.H. Ram\'{i}rez-Agudelo   \inst{1}
     \and
     H. Sana      \inst{2}
     \and
      S.E. de Mink \inst{1}
	\and
	V. H\'enault-Brunet \inst{3}
      \and
     A. de Koter  \inst{1,4}
      \and
  N. Langer \inst{5}
    \and
     F. Tramper \inst{1} 
	\and
      G.~Gr\"afener \inst{6}
       \and 
     C.J. Evans   \inst{7}
    \and     
      J.S. Vink  \inst{6}         
 \and
     P.L. Dufton    \inst{8}        
      \and
	W.D. Taylor  \inst{7}          
}
\institute{ 
          Astronomical Institute Anton Pannekoek, 
          Amsterdam University,  
          Science Park 904, 1098~XH, 
          Amsterdam, The Netherlands\\
          \email{o.h.ramirezagudelo@uva.nl}
\and 
           ESA/Space Telescope Science Institute,
           3700 San Martin Drive,
           Baltimore,
           MD 21218,
           USA 
\and 
    Department of Physics, 
    Faculty of Engineering and Physical Sciences, 
    University of Surrey, Guildford, 
    GU2 7XH, UK
\and 
           Instituut voor Sterrenkunde, 
           Universiteit Leuven, 
           Celestijnenlaan 200 D, 
           3001, Leuven, Belgium
\and 
           Argelander-Institut f\"ur Astronomie, 
           Universit\"at Bonn, 
           Auf dem H\"ugel 71, 
           53121 Bonn, Germany 
 \and 
           Armagh Observatory,
           College Hill,
           Armagh, BT61 9DG,
           Northern Ireland,
           UK 
\and 
           UK Astronomy Technology Centre,
           Royal Observatory Edinburgh,
           Blackford Hill, Edinburgh, EH9 3HJ, UK           
\and 
           Armagh Observatory,
           College Hill,
           Armagh, BT61 9DG,
           Northern Ireland,
           UK  
           }
             
   \date{Received ....}

 
 \abstract
  {  The initial distribution of spin rates of massive stars is a fingerprint of their elusive formation process. 
It also sets a key initial condition for stellar evolution and is thus an important ingredient in stellar 
population synthesis. So far, most studies have focused on single stars. Most O stars are however found in multiple systems.}
    {  
    By establishing the spin-rate distribution of a sizeable sample of O-type spectroscopic binaries 
and by comparing the distributions of binary sub-populations with one another as well as 
 with that of presumed single stars in  the same region, we aim to constrain the 
initial spin distribution of O stars in binaries, and to identify signatures of the physical mechanisms 
that affect the evolution of the massive stars spin rates. 
}
    { We use ground-based 
        optical spectroscopy obtained in the framework of the VLT-FLAMES Tarantula Survey (VFTS) to 
   establish the projected equatorial rotational velocities (\vrot) for components of 114 spectroscopic binaries in 30 Doradus. 
   The \vrot\ values are derived from the full-width at half-maximum (FWHM) of a set of spectral lines,   
   using a FWHM vs. \vrot\ calibration that we derive based on previous line analysis methods applied to single O-type stars in the VFTS sample. }
   { The overall \vrot\ distribution of the primary stars resembles that of single O-type stars in the VFTS,    featuring a low-velocity peak (at $\vrot < 200$\,\kms) and a shoulder at intermediate velocities ($200 < \vrot < 300$\,\kms).
    The  distributions of binaries and single stars however differ in two ways.
    First, the main peak at $\vrot \sim$100~\kms\ is broader and slightly shifted toward higher spin 
    rates in the binary distribution compared to that of the presumed-single stars. This shift
    is mostly due to short-period binaries ($P_\mathrm{orb} \lesssim 10$~d).
    Second,
    the \vrot\ distribution of primaries lacks a significant population of stars spinning faster than 300 \kms\ while
    such a population is clearly present in the single star sample.
    The \vrot\ distribution of binaries with amplitudes of radial velocity variation in the range of 20 to 200~\kms\ 
    (mostly binaries with $P_\mathrm{orb} \sim 10-1000$\,d  and/or with $q < 0.5$) is similar to that of single O stars 
    below $\vrot \lesssim 170$~\kms.}
    {Our results are compatible with the assumption that  binary components formed with the same spin distribution 
as  single stars and that this distribution contains few or no fast spinning stars.
 The larger average spin rate of stars in  short-period binaries  may either be explained by spin-up through 
tides in such tight binary systems, or by spin-down of a fraction of the presumed-single stars and long period binaries through magnetic braking (or by a combination of both mechanisms).
Most primaries and secondaries of SB2 systems with $P_\mathrm{orb} \lesssim 10\,$d appear to have similar rotational velocities. This is in agreement with tidal locking in close binaries of which the components
have similar radii. The lack of very rapidly spinning stars 
among binary systems supports the idea that most stars with $\vrot \gtrsim 300$~\kms\ in the single star sample
are actually spun-up post-binary interaction products.
Finally, the overall  similarities (low-velocity peak and intermediate velocity shoulder) of the spin distribution of binary and single stars
argue for a massive star formation process  in which the initial spin
is set independently of whether stars are formed as single stars or as components of a binary system.
}

   \keywords{
             stars: rotation -- 
             stars: binaries: spectroscopic -- 
             Magellanic Clouds --
             Galaxies: star clusters: individual: 30 Doradus --
			line: profiles             
               }

   \maketitle
%

\section{Introduction}
The spin rate that a massive star acquires at birth carries information about the  poorly constrained processes involved in its formation
\citep[e.g.,][]{zinnecker2007,rosen,2014arXiv1403.3417K}.  Once assembled, the spin of a massive star influences its nucleosynthesis and lifetime and more generally its evolutionary scenario \citep[e.g][]{2000ARA&A..38..143M,brott,ekstrom2012,2012A&A...542A..29G} and supernova characteristics \citep[e.g.,][]{yoon2005,woosley, 2013MNRAS.433.1114Y,2014A&A...564A..30G}.  For a review of these topics we refer to \citet{langer2012}.

Up to now, studies of the spin properties of massive stars have focused on the presumed-single star population
with only some reporting  measurements of spectroscopic binaries \citep[e.g.,][]{penny,penny2009}.
Here we present an analysis of the projected equatorial rotational velocity (\vrot) properties of the O-type spectroscopic binary population observed in the context of
the VLT-FLAMES Tarantula Survey \citep[VFTS;][]{evans}, a multi-epoch optical spectroscopic campaign
targeting 360 O-type objects and hundreds of cooler stars in the 30\,Doradus region in the
Large Magellanic Cloud.   Overviews of the VFTS project and the results obtained so far are provided in e.g., \citet{2011Msngr.145...33E}, \citet{dekoter1,2013ASPC..470..111D} and \citet{2013EAS....64..147S}. 

For the single O-type stars in the VFTS survey, \citet{ramirezagudelo} found a  \vrot\ distribution  rotational  peaking at about 80~\kms.  After deconvolution assuming arbitrary orientations of stellar spin vectors this peak shifts to about 100 \kms,
corresponding to $\sim$20\%\ of the break-up speed.  
Because spin-down through stellar winds is limited for the vast majority of the sample, \citet{ramirezagudelo} suggested that  this peak was representative of  the initial 
distribution of spin rates of single stars, i.e., of the outcome of the formation process. 


\citet{dufton} investigated the VFTS  spin distribution of non-supergiant early-B stars and found a strong bi-modality. The distribution presents a low-velocity peak at $\varv_\mathrm{e} < 100$~\kms\ and a broad and even stronger high-velocity component extending from $\varv_\mathrm{e} \approx 170$ to 350~\kms. The low- and high-rotation velocity regimes in the B-star sample are separated by a clear \vrot\ gap extending from 100 to about 170~\kms. The strong bi-modality of the spin-rate distribution seen in the VFTS B-star sample is generally not observed in  the VFTS single O-type stars \citep[][]{ramirezagudelo}. A hint of bi-modality may however be present in the O9 distribution and may suggest a smooth transition from the O star to the B star regimes. The weakly pronounced second O-star peak is however located  at a slightly lower projected spin velocity and is not as broad as the one seen in the B-star distribution. Further data would also be needed to validate the statistical significance of such a peak in the O9 star population.

According to the authors, the low spin velocity  peak in the B-star distribution may be the result
of magnetic spin-down, with long-lived magnetic fields possibly been generated through stellar mergers in close
binary systems. Since close binaries are unlikely to contain stellar merger products, such a
scenario would predict that the fraction of binary components with the lowest spin rates is reduced 
in the spin distribution of binary components.

The observed \vrot\ distribution of O stars further presents  a declining high-velocity tail at $\vrot\, \ga 300$\,\kms.
By comparison with the results of recent population synthesis  \citep{selma},  \citet{ramirezagudelo} 
suggested  that the high rotational velocity tail may be dominated by spun-up post-interaction objects. Such binary evolution products are 
indeed expected to be seen as single stars in radial velocity (RV) surveys due to the absence of large RV variations, either because of coalescence  of the binary components or because the mass donor has lost so much mass that the induced RV variations of the mass gainer -- that dominates the system optical brightness after mass transfer -- are below any practical detection limit for spectroscopic binaries \citep{selma2014a}. The mass gainer may also be truly single in cases where the binary is disrupted by the supernova explosion of the primary component.

The O-type binaries in the VFTS are identified by their  RV variations \citep{sana}
and hence will be dominated by pre-interaction systems \citep{selma2014a}. 
To first order, this sample  provides thus a view on the rotation properties of massive stars which is
unaffected by strong binary interactions, either through mass transfer or merging. In this respects, results from the binary sample may thus be confronted to that from presumed-single stars, where binary interaction effects are expected to lead to strong signals 
\citep{selma,selma2014a}. 

Furthermore, we may look at the role of tidal interactions
in systems with periods on the order of 1--10 days. This tends to synchronize the rotation of the components with 
the orbital motion, leading to a spin-up of stars that at first have modest rotation rates, and to the synchronization of the angular
velocities of primaries and secondaries in the same binary \citep{selma}.

Finally, we may take out the strong binary effects from the single star spin distribution,
that of tides from the binary component spin distribution, and then relate the
differences, if any, in the two resulting distributions to differences in the 
star formation processes of single and binary stars. 


%
This paper is organized as follows.
Section \ref{sec:sample} introduces our sample. 
Section \ref{sec:methodology} describes the adopted methodology to estimate the projected spin rates of the components of the VFTS O-type spectroscopic binaries.
Section \ref{sec:results} presents and analyses the obtained \vrot\ distributions. 
Our results are discussed in Section \ref{sec:discussion} and our conclusions are summarized in 
Section \ref{sec:conclusions}. 


\section{VFTS data and O-type binary sample}\label{sec:sample}

The VFTS campaign has been described in \citet{evans}.  All spectra discussed here were obtained with the Medusa fibre-feed to the Giraffe spectrograph. The VFTS Medusa sample contains 332 O-type objects. Spectral classification and RV measurements were presented in \citet{walborn2014}
and  \citet{sana}, respectively. 

The method adopted for RV measurements  was based on Gaussian fitting and, for each O star in the sample, provided Doppler shift, amplitude and full width at half maximum (FWHM) values for a subset of  \ion{He}{i}\ and \ion{He}{ii}\ lines.
116 objects were classified as spectroscopic binaries (SB) on the basis of their statistically significant and large-amplitude RV variations ($\rm{\Delta}\,RV\,  >$ 20\,\kms). 

\begin{figure}
\centering
\includegraphics[scale=0.38]{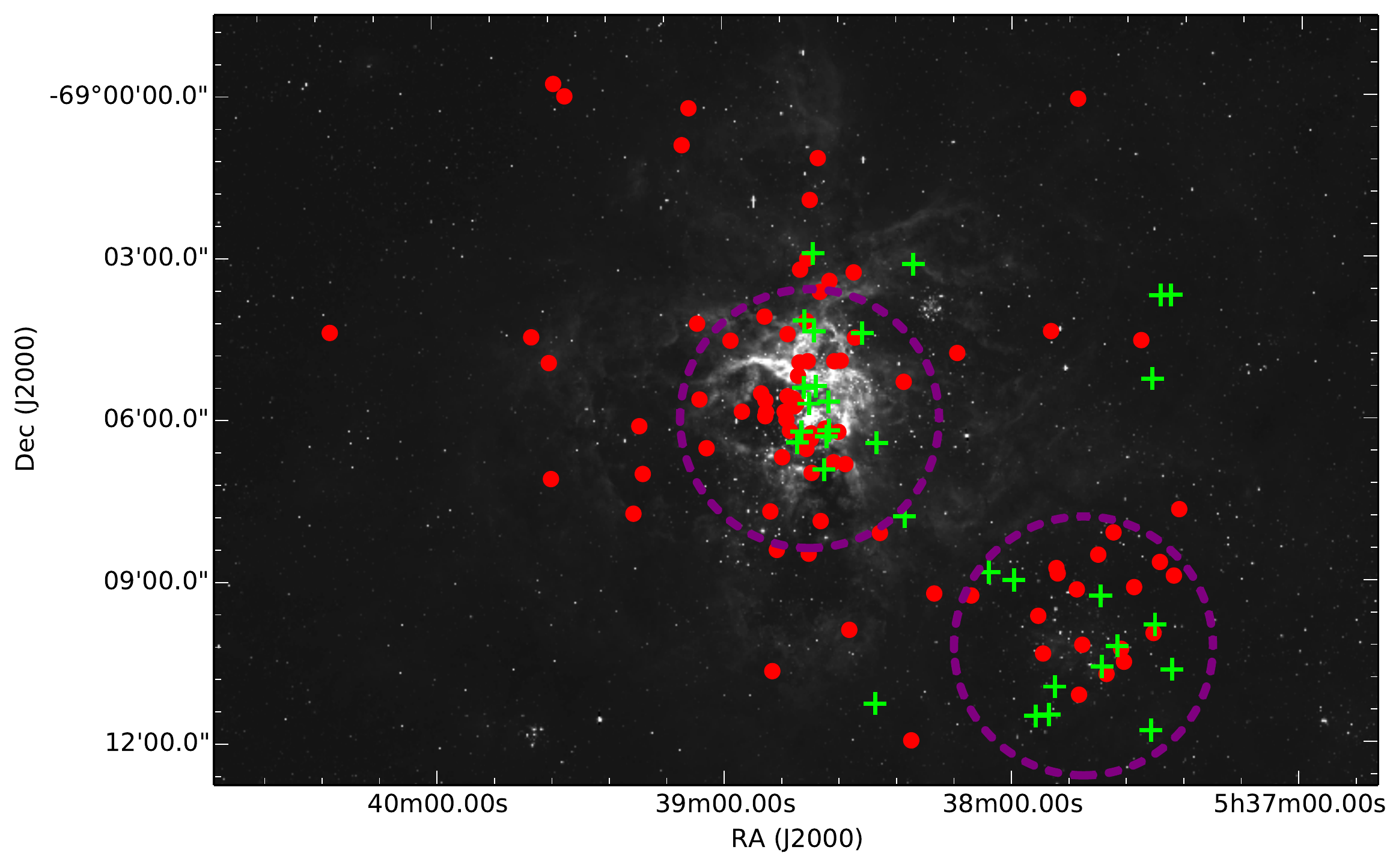}
\caption{Spatial distribution of the O-type binary systems in the 30\, Dor field of view.
                Dots indicate SB1 systems (85 objects) and crosses SB2 systems (31 objects). 
                The dashed circles  have radii of 2.4\arcmin\ and are centered on the central 
                cluster NGC\,2070 and on NGC\,2060, at 6.7' to the 
                south-west of NGC\,2070.
                (A color version is available in the online version of the paper).  }
\label{fig:field_binaries}
\end{figure}

The identified O-type binary sample contains 85 single-lined (SB1)  and 31 double-lined  (SB2) spectroscopic binary systems. Figure~\ref{fig:field_binaries} shows the spatial distribution of the binary stars.  The systems are concentrated in the two clusters, NGC\,2070 and NGC\,2060 (6.7' south-west of NGC\,2070), although a sizeable fraction is distributed
 throughout the field of view.

\begin{figure*}
\centering
\includegraphics[scale=0.5]{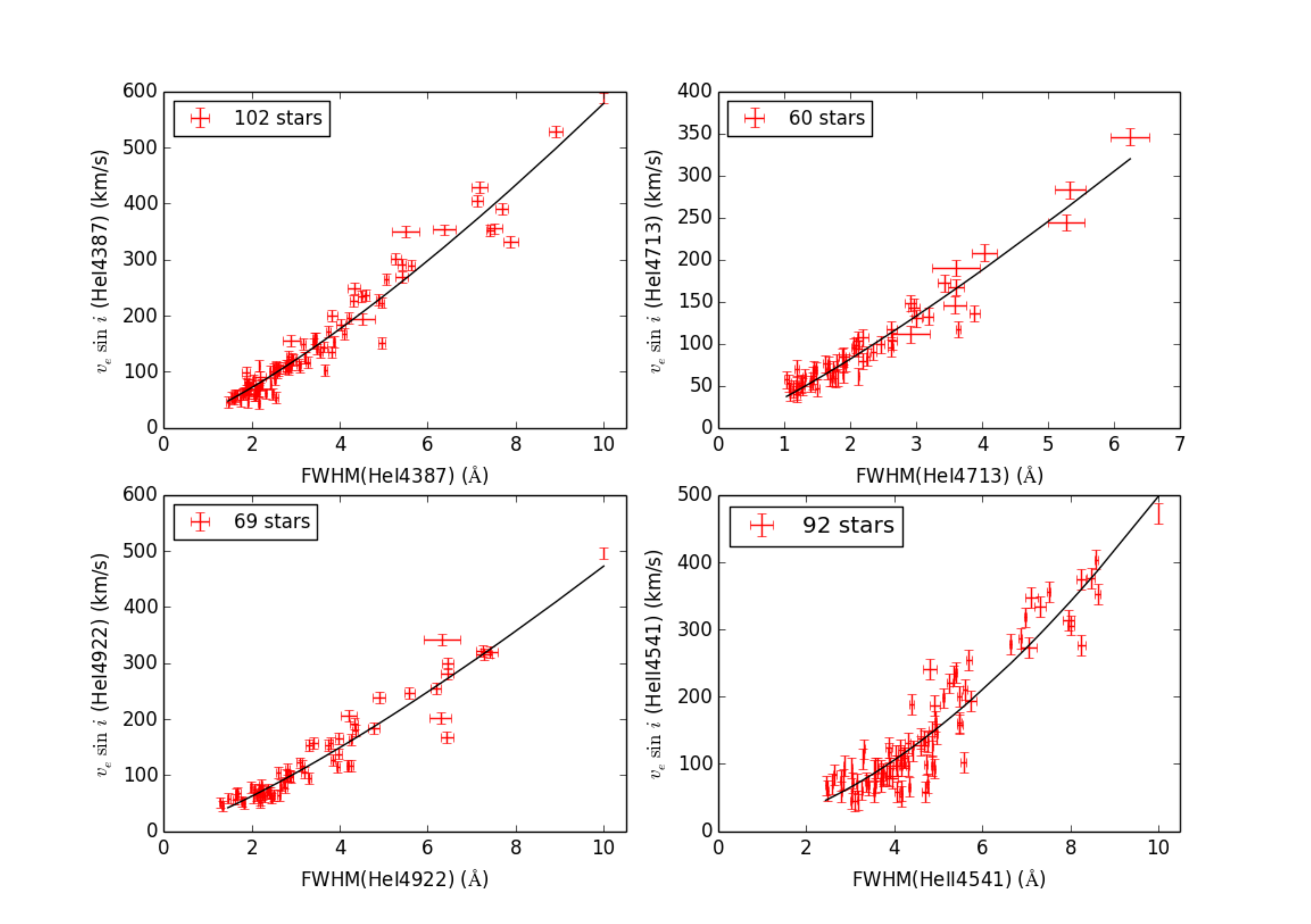}
\caption{\vrot\ derived by means of Fourier transform or line-profile fitting methods 
(vertical axis; see \citeauthor{ramirezagudelo} \citeyear{ramirezagudelo})
versus FWHM for \hel{i}{4387}, 4713, 4922, and \hel{ii}{4541} (horizontal axis). 
The lines show the fitted FWHM $-$ \vrot\ calibration relations described by Eq.~\ref{eq:1}
and Table~\ref{table:1}.
}
\label{fig:vsini_fwhm}
\end{figure*}

\section{Projected rotational velocities}\label{sec:methodology}

The projected rotational velocity, \vrot, of stars
can be measured from the broadening of their spectral
lines \citep[][]{carroll,Gray}. 
For OB-type stars commonly used methods include the direct measurement of the full width at half maximum 
\citep[e.g.,][]{slettebak,herrero,abt},
cross-correlation of the observed spectrum against a template
spectrum \citep[e.g.,][]{penny,howarth}, and comparison with synthetic lines 
calculated from model atmospheres \citep[e.g.,][]{mokiem}.
\citet{simon,simon2014} apply a 
Fourier transform (FT) and a line-profile fitting method.

The phase coverage of the VFTS binaries does not allow us to disentangle the spectra of the binary components, hence one cannot easily apply FT or line profile fitting methods. To obtain the projected rotational velocities of the binary sample in a homogeneous way, we thus focus on the FWHM of specific spectral lines. 
The FWHMs are indeed easily quantifiable and less affected by line blending and/or the presence of faint or undetected companions than more sophisticated  methods. The drawbacks are relatively large error bars on individual measurements (see below) and a limited discriminative power at $\vrot \lesssim 100$~\kms\ due to the fact that the method does not disentangle rotational broadening from other types of broadening  such as macro-turbulence and Stark broadening.

%


\subsection{FWHM $-$ \vrot\ calibration}\label{subsec:calibration}



For the single O-type stars in the VFTS sample, we used  the projected rotational velocities  previously obtained using 
 both a FT and a line profile fitting method \citep{ramirezagudelo}. 
The FT method is able to disentangle rotational broadening 
from other broadening agents (such as for instance broadening due to macro-turbulent velocity fields) in cases
where rotational effects dominate.  For stars where rotation does not dominate the broadening, the line fitting method also 
takes into consideration the presence of additional broadening \citep[for more details see][]{ramirezagudelo}.

The  \hel{i}{4387}, 4922, 4713, and \hel{ii}{4541} lines studied in \citet{ramirezagudelo} have also been used for RV measurements 
\citep{sana} and thus have both \vrot\ and FWHM measurements available.
This allows for a calibration of the FWHM values in terms of \vrot. 

Figure~\ref{fig:vsini_fwhm} compares \vrot\ vs. FWHM for the helium lines listed above.
The number of stars for which this comparison can be made differs per spectral line 
and depends on spectral type and data quality.
The main reason for this is that the lines need to pass a set of quality criteria to ensure reliable measurements, 
i.e.,  they should be of sufficient
strength relative to the signal-to-noise ratio of the spectrum and to have a \vrot\ measurement that is above the
resolution limit \citep{ramirezagudelo}.  

All four lines show a clear trend between \vrot\ and FWHM. 
The scatter in the relation for \hhel{ii} is considerably larger at relatively low \vrot\ than for the three \hhel{i} lines.  
This results from the fact that the Stark broadening  is more pronounced in \hhel{ii} lines , and that a treatment of 
this broadening is not properly included in \citet{ramirezagudelo}.

To calibrate FWHM as a function of \vrot\ we obtain a phenomenological description of the trends seen in  Fig.~\ref{fig:vsini_fwhm}. We successively adjusted polynomials of order 1, 2 and 3 as well as a power-law function with and
without an offset. We  perform F-tests to decide whether additional fit parameters result in significantly better  representations of the data. 
All  fit functions, except the first order polynomial, give similar quality results.  As the power law without an offset has the fewest free parameters, we opted for this representation\footnote{While an extra offset would formally express the fact that spectral lines have an intrinsic width -- hence would be more physical -- adding this extra parameter yields no improvements on the fit quality and we opted for the simpler function.}, i.e.,
%
\begin{equation}
     \vrot = a \times {\rm FWHM}^{b},
\label{eq:1}
\end{equation}
where $a$ and $b$ are the fitting parameters.  

Table~\ref{table:1} summarizes the best fit values and their uncertainties for each of the lines. 
The root mean square (rms) residuals 
are about 15 to 25\,\kms\ for the \hhel{i} lines and reach 35\,\kms\ for the \hel{ii}{4541} line.
In the next sub-section, we lay out our strategy in using these four relations and investigate the uncertainty in the 
\vrot\ determinations that follow from this approach.

\begin{table}
\caption{
FWHM $-$ \vrot\ calibration based on the VFTS single O-type star sample. The columns list the
diagnostic line (column 1), the number of stars for which the line fulfills the selection criteria (column 2), the coefficients of fit function Eq.~\ref{eq:1} 
(columns 3 and 4) and the root mean square (rms) deviation between fit and data in \kms\ (column 5).
}              
\label{table:1}      
\centering                          
\begin{tabular}{lrccc}
\hline
\hline
Diagnostic line    & \# stars         	  & $a$     &  $b$ & rms     \\  
\hline
                   &                      & (\kms/\AA) & & (\kms) \\
\hline
\hel{i}{4387}  &  102 & 28.99 $\pm$ 1.55    &    1.30 $\pm$ 0.03  &    25       \\  
\hel{i}{4713}  &    60 & 35.51 $\pm$ 1.81    &    1.20 $\pm$ 0.04  &    15       \\  
\hel{i}{4922}  &    69 & 26.00 $\pm$ 2.01     &    1.26  $\pm$ 0.04 &   25       \\  
\hel{ii}{4541} &    92 & 10.17 $\pm$ 1.27    &    1.69 $\pm$ 0.07  &    35      \\  \hline
\end{tabular}
\end{table}

\subsection{FWHM$-$based \vrot\ values for the single star sample}
\label{sec:comparison_single_sample}

We first apply the FWHM$-$\vrot\ calibration to the sample of 216 presumably single O-type stars, 
as this provides an internal consistency test.  For four stars (VFTS\,400, 444, 565, and 587), 
no FWHM measurements of any of the four diagnostic lines are available.
These have been excluded from our comparison, leaving 212 stars. 

\begin{figure}[t!]
\centering

\vspace{-8pt}

\includegraphics[scale=0.45]{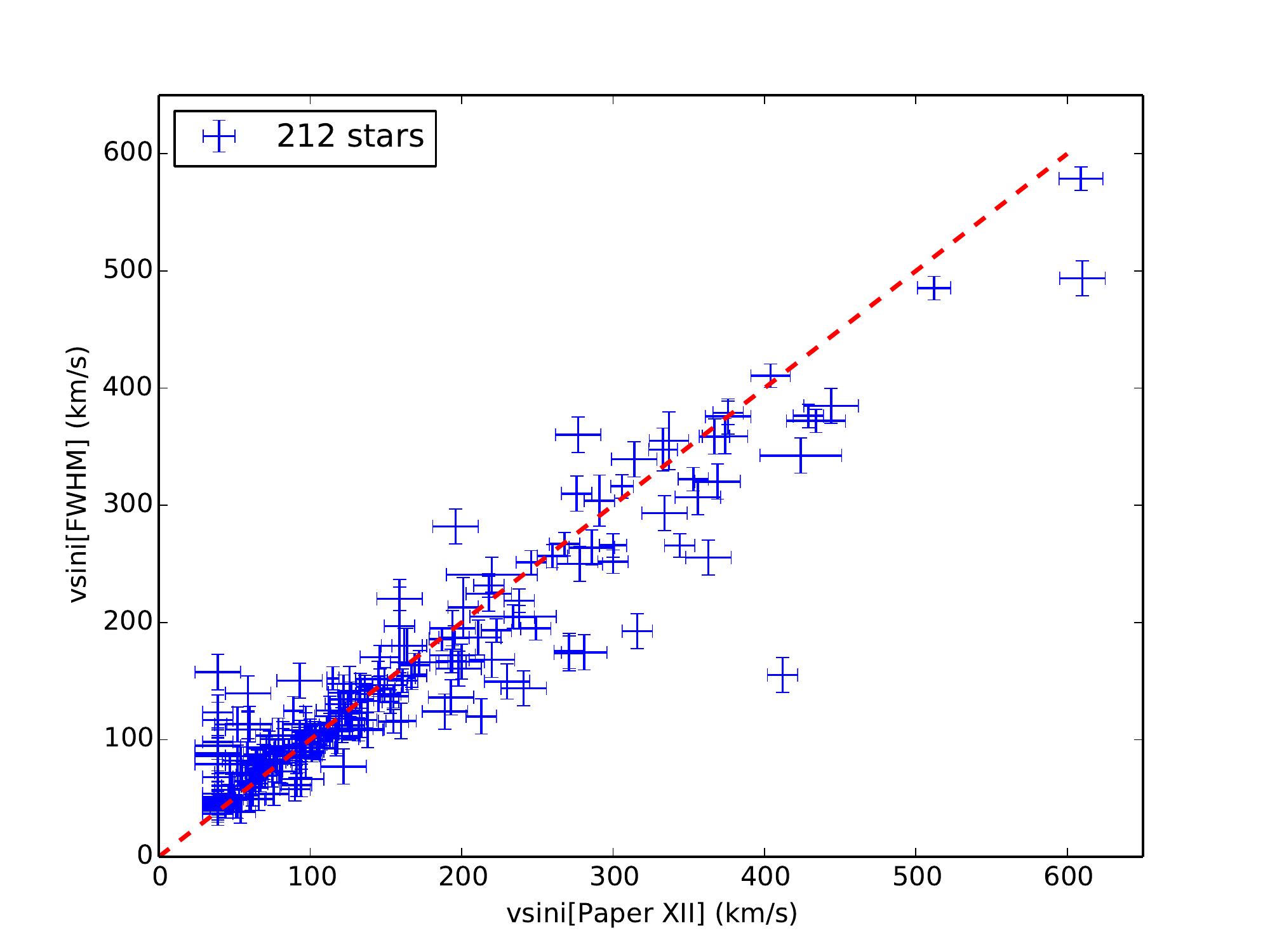}
\caption{Comparison of  \vrot\ values derived from the FWHM $-$ \vrot\ calibration (Eq.~\ref{eq:1}) and those of \citet{ramirezagudelo},
		which are based on more sophisticated methods. The dashed line shows the one-to-one relation.}
\label{fig:comparison_paper}
\end{figure}

In \citet{ramirezagudelo}
the diagnostic lines have been divided into two groups, one of relatively good quality lines (Group A)  and one of lesser
quality lines (Group B).  The Group A lines include the \hhel{i} lines at 4387, 4713, and 4922 \AA.   Only if these lines
are very weak or not present,  \vrot\ relied on \hel{ii}{4541}. These latter measurements are referred to as Group B.
We also adopt this division of measurements 
in two groups in order to decide which lines provide the most reliable value of \vrot.  If more than one \hhel{i}
line is available in a Group A source, we use the unweighted average to obtain the mean projected
rotational velocity.  As this mean is based on at most three values, the standard deviation tends to underestimate the true dispersion.
We thus adopt the  rms residuals of Table~\ref{table:1}
 as error bars for our measurements, unless the standard deviation of a given star is larger.  In that
case we adopt the computed value. 

Figure~\ref{fig:comparison_paper} compares the \vrot\ values based on the FWHM\-$-$ \vrot\ calibration  with those
obtained in \citet{ramirezagudelo}.  The systematic difference is less than 5\,\kms\ and the standard deviation
is about 40\,\kms. 
For slowly spinning stars, the calibration presented
here tends to somewhat overestimate \vrot.  
Indeed in this low-velocity regime the FWHM$-$\vrot\ calibration fails to disentangle the contribution of rotational broadening 
from other  broadening mechanisms (see Sect.~\ref{subsec:calibration}). At large velocity (\vrot\, $>$ 400~\kms), our FWHM-based measurements tend to underestimate \vrot\ compared to values obtained with the FT or profile fitting method. It may suggest the presence of a higher-degree component in the 
calibrations but our previous tests show that, if present, the latter could not be firmly identified
considering our data quality.

\begin{figure}
\centering
\includegraphics[scale=0.4]{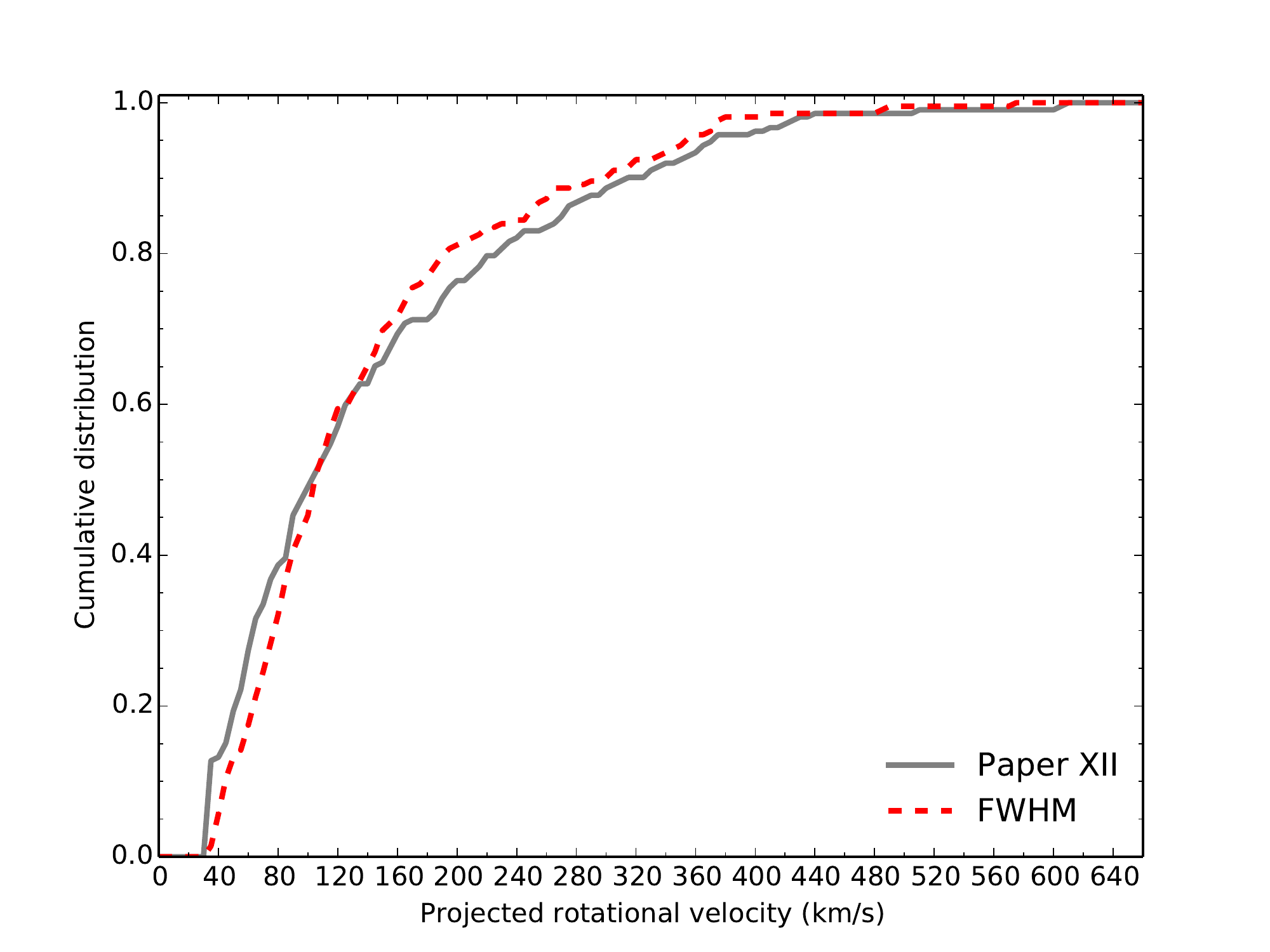}
\includegraphics[scale=0.4]{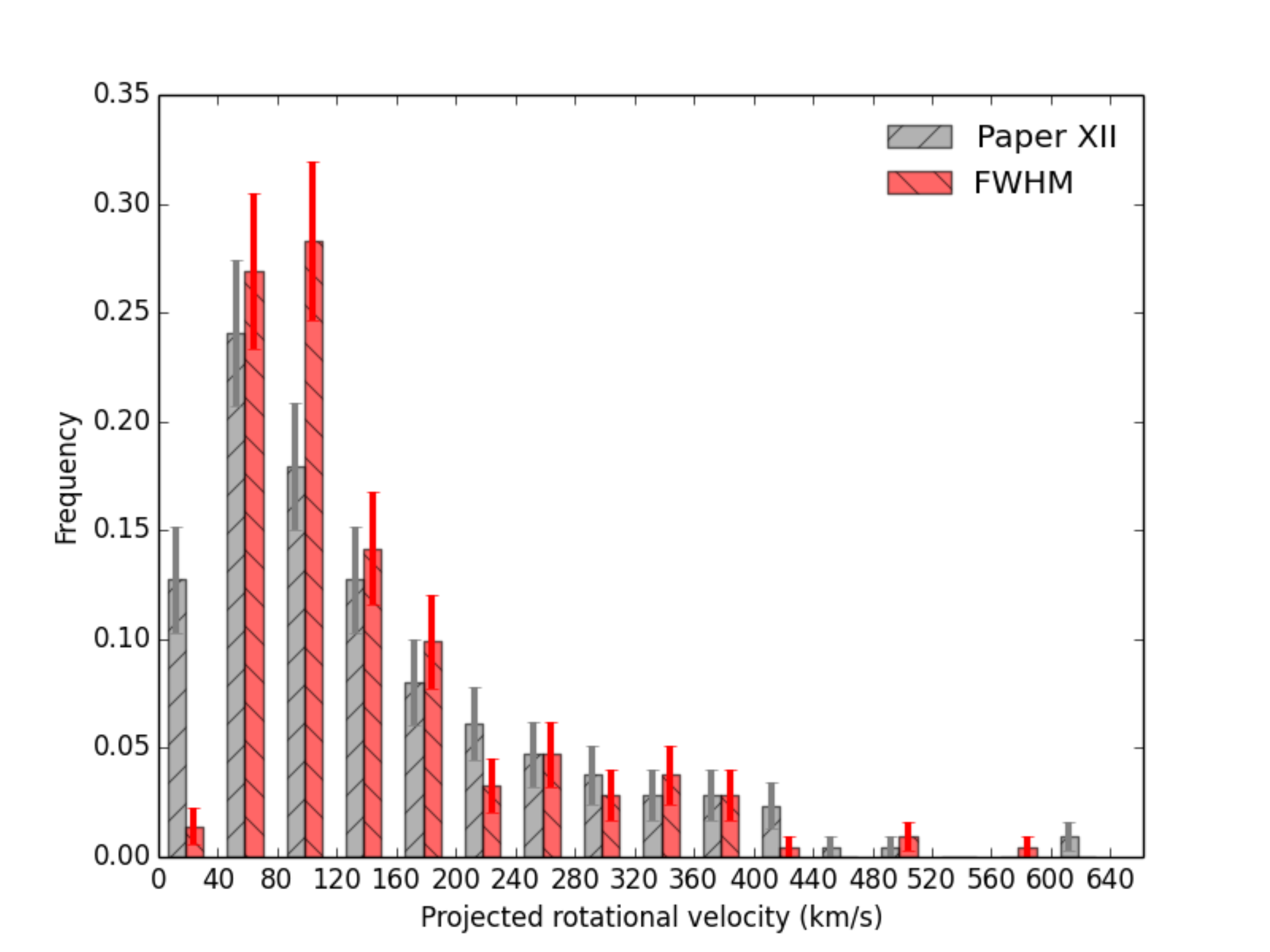}
\caption{
Cumulative (upper panel) and frequency (lower panel, with Poisson error bars)
distributions of the projected rotational velocities of the presumed-single O-type stars obtained 
using the FWHM$-$\vrot\ calibration (Eq.~\ref{eq:1}) and those of \citet{ramirezagudelo}.
}
\label{fig:hist_comparison_paper}
\end{figure}

Figure~\ref{fig:hist_comparison_paper} compares the \vrot\ probability density distribution (top panel) and \vrot\ frequency
distribution (bottom panel) of \citet{ramirezagudelo}
with that derived from the FWHM$-$\vrot\ calibration, adopting a bin size of 40\, \kms\ as in \citet{ramirezagudelo}.
The first bin of the new calibration is depopulated relative to the \citet{ramirezagudelo} analysis because of the
neglect of
additional broadening processes.  Such extra broadening may affect \vrot\ for stars rotating below 150
\kms, and also explains the shift of the peak of the distribution to somewhat higher \vrot\ values.  At velocities above 200 \kms,
the distribution is not affected in a statistically significant way, and the small (one bin) changes that occur can be
explained by the random uncertainties in the two methods. 

 Importantly, the general shape of the distribution --
a peak at about 100 \kms\ and a high-velocity tail -- is preserved using the FWHM$-$\vrot\ method. Despite relatively large individual measurement errors, this method still allows us to statistically investigate the rotational properties of massive stars, including those of binaries for which more sophisticated methods are more cumbersome to apply.

\section{The \vrot\ distribution of spectroscopic binaries} \label{sec:results} 

In this section, we construct and investigate the \vrot\ distribution of stars in O-type binary systems.  
Out of 116 primaries \vrot\ values could not be established  in two cases, VFTS~531 and VFTS~810, as the spectra do not pass the quality criteria described in Sect.~\ref{subsec:calibration} and thus
these two stars were excluded from our analysis.

From here on, we only use \vrot\ values obtained using the FWHM$-$\vrot\ calibration, i.e.,  also for the set of single stars. This allows for a direct and consistent comparison of the single and binary samples. 

We further perform statistical comparisons of the \vrot\ distribution from different sub-samples 
in order to identify statistically significant signals in our data.
We use  Kuiper's test of hypothesis \citep[KP,][]{kuiper} to search for significant differences between the 
parent populations of the considered samples. 
Results are reported in Table~\ref{table:2}, where we list the sum $D$ of the absolute values of
the most positive and most negative differences between the two cumulative distribution functions that
are being compared and the Kuiper probability $p_{\rm K}$. 
Specifically, KP tests allow us to test the null hypothesis that two observed distributions are randomly drawn
from the same parent population. 
This test is similar to the more well-known Kolmogorov-Smirnov test but presents the 
advantage that it is as sensitive to variations in the tail as it is to variations in the middle of the distributions.

In Table~\ref{table:2new} we also present the frequency of stars in a given sub-population that are 
spinning faster than the specified limit. We will use these statistics to assess the significance of the presence of a high-velocity tail in these sub-populations.


\begin{table}
\caption{Kuiper's test statistics. The figure index is given in the first column.  
The considered samples and number of stars in each of these samples are listed in Cols. 2 to 5.  
Column 6 specifies the KP test statistic $D$ and 
column 7 the probability $p_{\rm K}$. 
}             
\label{table:2}      
\centering                          
\begin{tabular}{lllrrrr}        
\hline\hline\\[-8pt]                 
Fig. & Sample 1 & Sample 2 &  n1 & n2&  D   & $p_{\rm K}$  \\[1pt]     
       &  &  &   &  &  & (\%) \\ \hline \\[-9pt]     
\ref{fig:dist_single_primaries} & Presumed-single                &   Primaries              &  212    & 114 & 0.20 & 3    \\     
\ref{fig:dist_single_secondaries} & Primaries          &   Secondaries         &  114    &    31 & 0.18 &  95    \\     
\ref{fig:dist_binaries}& Low-$\rm{\Delta RV}$ & High-$\rm{\Delta RV}$      & 85 & 29   & 0.24  & 57 \\
\ref{fig:prim_sec}                          & Primaries (SB2)          &   Secondaries (SB2)           &      31    &    31 & 0.16 &  99   \\     \hline\\[-9pt]
 \multicolumn{7}{c}{For \vrot\ $\leq$ 170\,\kms}\\[1pt]
\ref{fig:dist_single_rvlow}           & Presumed-single        &   Low-$\rm{\Delta RV}$    &     156    &   64 & 0.15 &  76   \\

\hline                        
\end{tabular}
\end{table}

\begin{table}
\caption{Frequency of stars ($f$) from different sub-samples that are spinning more rapidly than the specified limit.
68\% confidence intervals (approx. $\pm$ 1 sigma error bars) were computed following binomial statistics \citep{2014ApJS..215...15S}.}             
\label{table:2new}      
\centering                          
\begin{tabular}{lccl}        
\hline\hline\\[-8pt]                 
     &  \multicolumn{3}{c}{ $f(\vrot)$}  \\[1pt] \\[-8pt]    
 Sample                                        & $>$ 200\,\kms         & $>$ 250\,\kms     &  $>$ 300\,\kms  \\[1pt]  \hline\\[-8pt]    
Presumed-single                        & 0.193 $\pm$ 0.028 & 0.156 $\pm$ 0.024 & 0.104 $\pm$ 0.019 \\
Primaries                                      & 0.175 $\pm$ 0.035 & 0.088 $\pm$ 0.026 & 0.026 $\pm$ 0.018 \\  
Primaries (SB2)                          & 0.161 $\pm$ 0.065 & 0.032 $\pm$ 0.032  & 0.000 ($<$ 0.032) \\
Secondaries                                & 0.097 $\pm$ 0.065 & 0.032 $\pm$ 0.032  & 0.032 $\pm$ 0.032 \\
Low-$\rm{\Delta RV}$               & 0.141 $\pm$ 0.035 & 0.082 $\pm$ 0.035 &  0.035 $\pm$ 0.024 \\
High-$\rm{\Delta RV}$              & 0.276 $\pm$ 0.069 & 0.103 $\pm$ 0.069 &  0.000 ($<$ 0.034) \\
\hline                        
\end{tabular}
\end{table}

\subsection{Primaries vs. single stars} \label{subsec:primaries} 


\subsubsection{Observed distributions}
Figure~\ref{fig:dist_single_primaries} compares the \vrot\ distributions of presumed-single stars and primaries. 
There is a surprising overall qualitative similarity between the two distributions.
Both are dominated by a low-velocity peak ($<$ 200~\kms) which contains $\sim$80\% of the samples, and 
show a shoulder at intermediate velocities ($200~\kms \leq \vrot \leq 300$~\kms). 
The distribution 
of single stars also features a clearly identifiable high-velocity ($>$ 300~\kms) tail, which is hardly seen in the distribution of primaries.

Though qualitatively similar the main low-velocity peaks are not identical.
The main peak in the distribution of the primaries is wider than that of the presumed-single sample and
overpopulates the distribution in the region between 100 and 300~\kms\  by 18\%. 
At large projected spin rates ($\vrot > 300$~\kms),
there is a deficiency of very fast rotating
primaries with respect to the presumed-single stars. The presumed-single sample harbours 22 stars out of 212  that exhibit 
projected rotational velocities larger than 300\, \kms\ ($10.4\pm1.9$\,\% of that sample) while the sample of primaries  only has 3 stars out of 114 
($2.6\pm1.8$\,\%, see Table~\ref{table:2new}). 
A KP test confirms  that  parent \vrot\ distributions  of primaries and of single stars are statistically different ($p_K = 3$\,\%).


\begin{table}[t!]
\centering
\caption{Baysian estimates of the rotation distribution parameters of Eq.~\ref{pdf_ve}. Reported values correspond to the 50th percentiles and confidence interval to the  16th and 84th percentiles of the posterior distributions.}
\label{tab: bayes}
\begin{tabular}{c c c}
\hline
\hline
Parameters & Primaries & Single stars \\
\hline
\multicolumn{3}{c}{Adjusted parameters}\\
\hline
\vspace*{-2mm}\\
$\alpha$ & $  7.83^{+ 1.39 }_{ -1.55 }$  & $  8.94^{+0.77  }_{-1.33  }$\\
\vspace*{-2mm}\\
$\beta$  & $0.0476^{+0.0091}_{-0.0101}$  & $0.0729^{+0.0086}_{-0.0121}$\\
\vspace*{-2mm}\\
$\mu$    & $   215^{+133   }_{-111   }$  & $   320^{+40    }_{-40    }$\\
\vspace*{-2mm}\\
$\sigma$ & $   142^{+171   }_{ -78   }$  & $   113^{+26    }_{-23    }$\\
\vspace*{-2mm}\\
$I_\gamma$ & $ 0.94^{+0.04 }_{-0.15 }$  & $ 0.71^{+0.06 }_{-0.07 }$\\
\vspace*{-2mm}\\
\hline
\multicolumn{3}{c}{Computed parameters}\\
\hline
\vspace*{-2mm}\\
   $\mu_\gamma=\alpha/\beta$          & $164^{+11}_{-11}$ & $122^{+8}_{-7}$\\
\vspace*{-2mm}\\
$\sigma_\gamma=\sqrt{\alpha/\beta^2}$ & $ 59^{+9}_{-6}$ & $ 41^{+5}_{-3}$\\

\vspace*{-2mm}\\
\hline
\end{tabular}
\end{table}

\subsubsection{Impact of inclination distributions}
\begin{figure}
\centering
\includegraphics[scale=0.4]{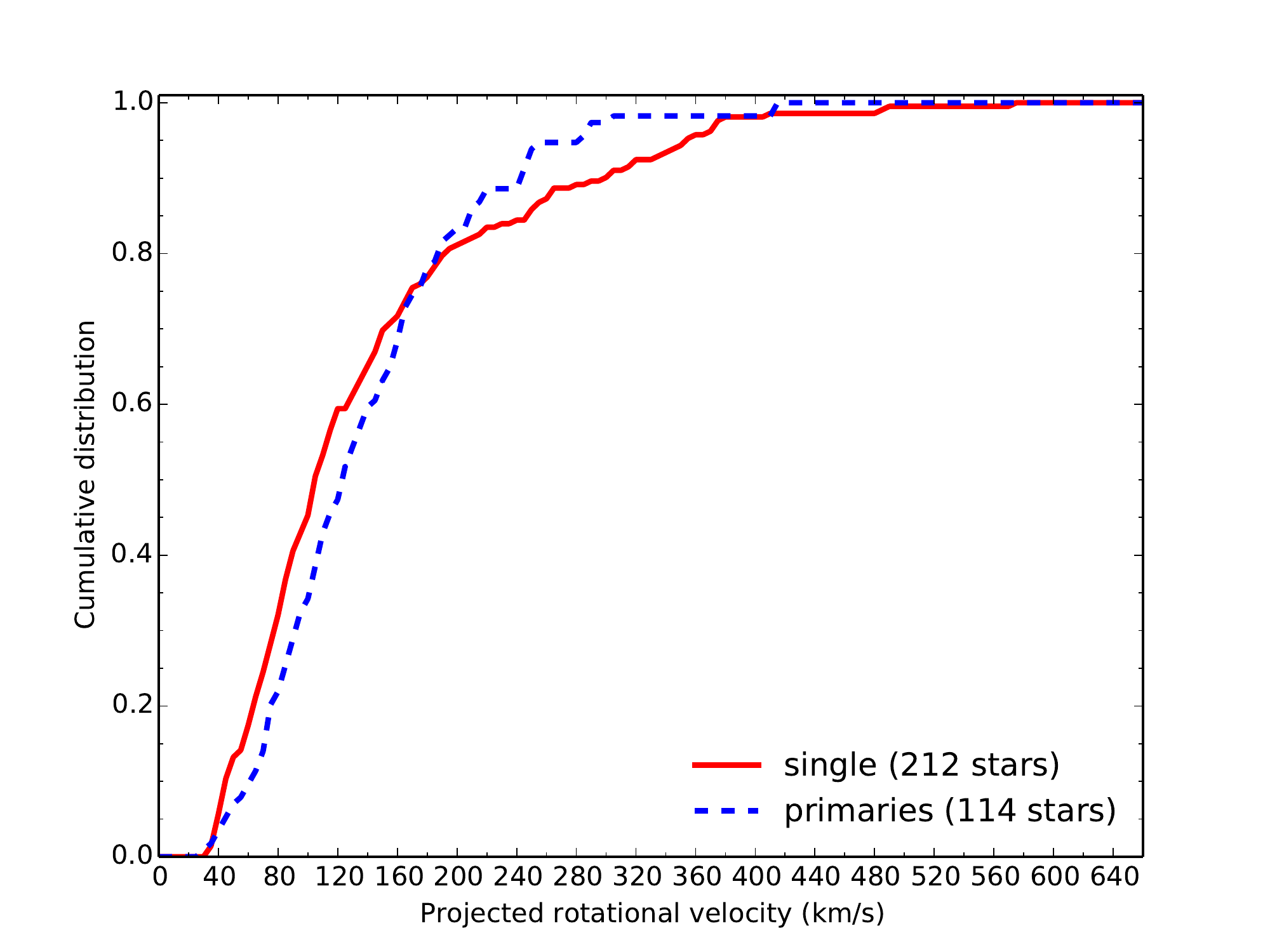}
\includegraphics[scale=0.4]{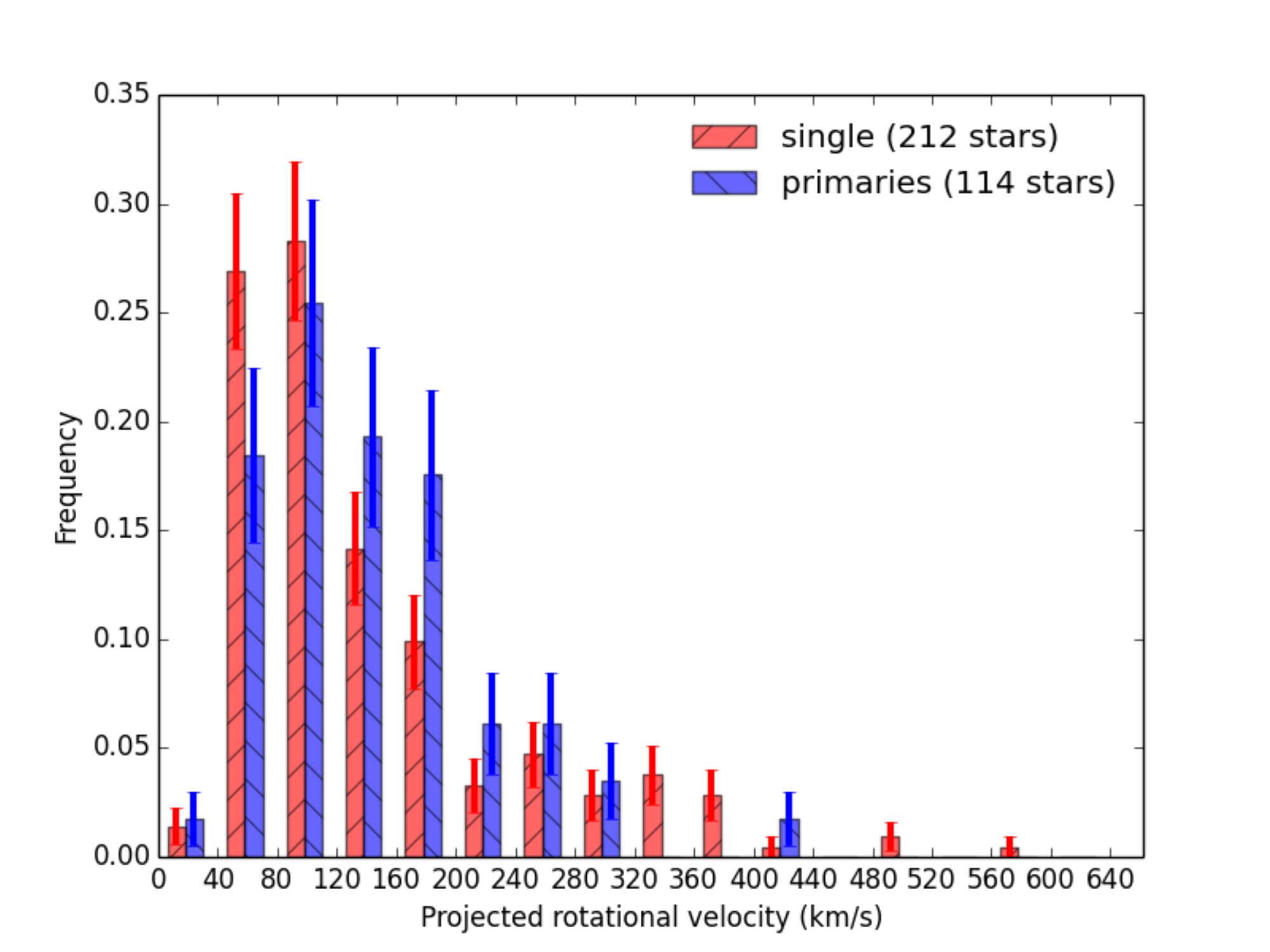}
\caption{Cumulative (upper panel) and frequency (lower panel, with Poisson error bars)
distributions of the projected rotational velocities of the presumed-single O-type stars (\vrot\,$-$ FWHM method)
and O-type primaries.
($\rm{\Delta RV}\, \leq$ 20~\kms\,) 
}
\label{fig:dist_single_primaries}
\end{figure}

Binary systems that are seen closer to edge-on, i.e., that have a large inclination angle, are more easily identified as binaries because of larger projected radial velocity variations. This may introduce an observational bias in the spin 
rate distribution of binaries relative to that of single stars if the spin axis of the components is preferentially oriented 
along the orbital rotation axis. For identical intrinsic spin distributions,  the observed distribution of single stars might thus be shifted to lower \vrot\ than that of binaries, potentially explaining part of the shift that we see in Fig.~\ref{fig:dist_single_primaries}. 

An estimate of the most extreme effect of this bias is provided by comparing the deconvolved $\varv_{\rm e}$ distribution of single stars \citep[fig. 16 of][]{ramirezagudelo} to the observed \vrot\, distribution of binaries in Fig.~\ref{fig:dist_single_primaries}. The latter indeed corresponds to the intrinsic distribution of binaries under maximum detection likelihood corresponding to $i=90$\degr\ for all systems. Even in these conditions, the main peak of the binary distribution is still broader and shifted towards higher $\varv_\mathrm{e}$ compared to the main peak of the single star distribution. Furthermore, we show in Appendix~\ref{app: bayes} that the actual distribution of inclination angle of the binaries in our sample is not sufficiently different to that expected from random orientation in 3D space to significantly contribute to differences seen in Fig.~\ref{fig:dist_single_primaries}.  We conclude that the possible bias induced by a preferred  binary detection for higher inclination system cannot be invoked to explain the difference of the $\varv_\mathrm{e}$ distributions in the primaries and presumed-single star samples. 

\subsubsection{Intrinsic $\varv_\mathrm{e}$ distributions}
In Appendix~\ref{app: bayes}, we use a simulated distribution of orbital inclination of the binaries in our sample and a Bayesian method to compute the intrinsic distribution of rotational velocities of the primaries in our sample. We adopt a general analytical form of the rotation velocity probability density function $f(\mathrm{\varv_\mathrm{e}})$,  similar to the one found for single O stars within VFTS \citep{ramirezagudelo}, i.e., a combination of a gamma  distribution and of a normal distribution:

\begin{equation}
f(\mathrm{\varv_\mathrm{e}}) = I_\gamma\ g(\mathrm{\varv_\mathrm{e}}; \alpha,\beta) + (1- I_\gamma) \ N(\mathrm{\varv_\mathrm{e}}; \mu, \sigma^2)
\label{pdf_ve}
\end{equation}
where
\begin{eqnarray}
g(x; \alpha,\beta) &=& \frac{\beta^\alpha}{\Gamma(\alpha)} x^{\alpha-1} e^{- \beta x}, \label{pdf_ve1}\\
%
N(x; \mu, \sigma^2) &=& \frac{1}  { 1+\mathrm{erf}(\mu /\sqrt{2}\sigma)} \frac{e^{-(x-\mu)^2 / 2 \sigma^2} }{\sqrt{2 \pi} \sigma} \label{pdf_ve2}.
\end{eqnarray}
In Eq.~\ref{pdf_ve2}, $\mathrm{erf}$ is the error function. It is used to renormalize the Gaussian component to unity between $x=0$ and $\infty$. Similarly, $I_\gamma$ and $(1- I_\gamma)$ are the relative contributions of both distributions to $f(v_\mathrm{e})$, such that $f(v_\mathrm{e})$ is also normalized to unity when integrated over positive $\varv_\mathrm{e}$. Such renormalizations were not implemented in the same way in \citet{ramirezagudelo}. For consistency in the comparison, we thus also applied  the Bayesian analysis method to the single star \vrot\ measurements obtained through our FWHM calibration curves.

\begin{figure}
\centering
\includegraphics[width=\columnwidth]{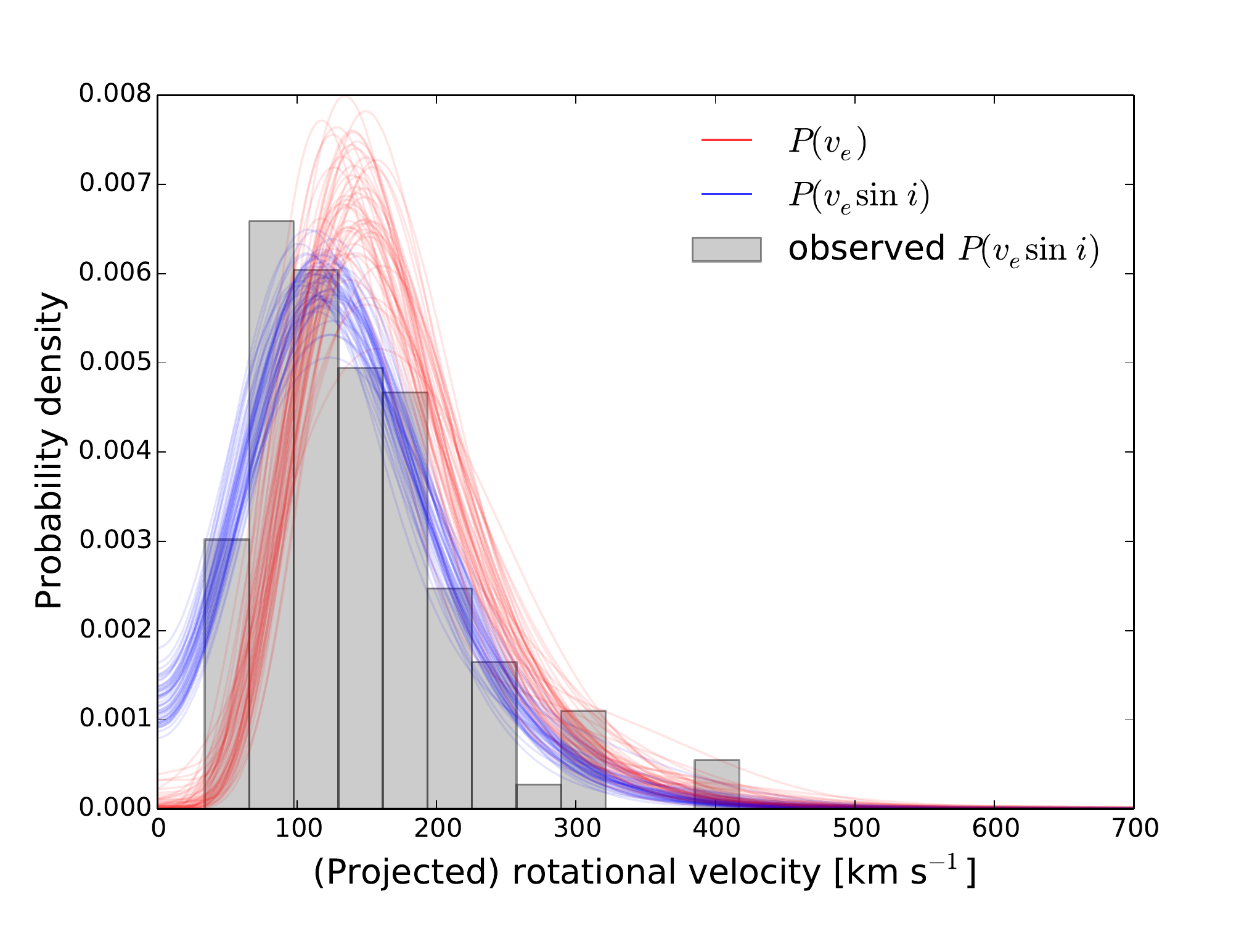}
\includegraphics[width=\columnwidth]{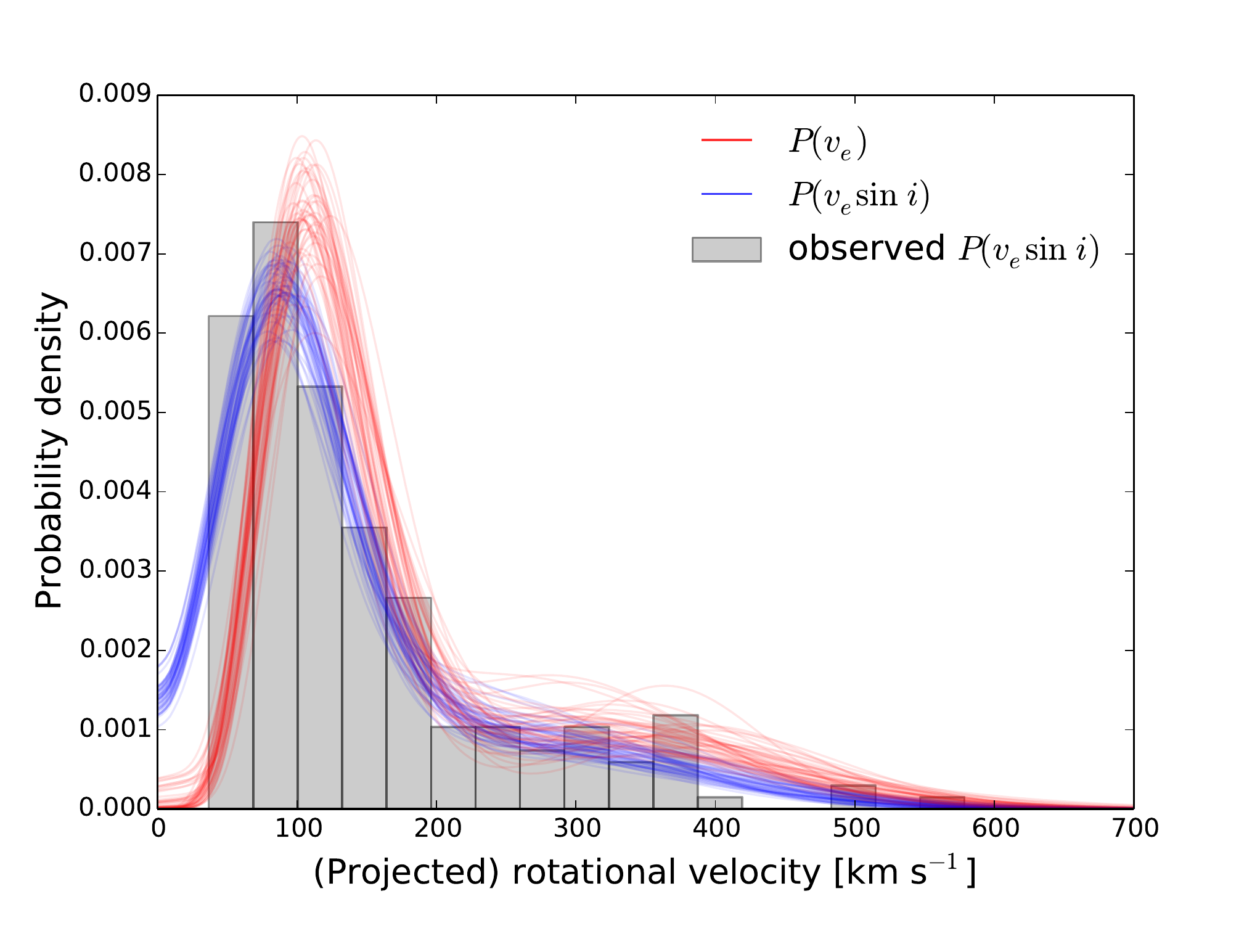}
\caption{Rotational velocity distributions for 50 random samples of the MCMC chain for the primaries ({\it upper panel}) and the single stars  ({\it lower panel}). The intrinsic rotational velocity distributions are shown in red.  In blue we show the corresponding distributions projected (assuming the $\sin{i}$ distribution inferred for the detected binaries in VFTS for the primaries and a random orientation of the system in 3D space for the single stars) and convolved with a Gaussian error distribution with $\sigma_v = 20 \ \kms$. Although the observed data is represented here by a histogram, the fit to the data is performed discretely, i.e., without any binning. The histogram bins are chosen according to Knuth's rule \citep{knuth2006}.}
\label{fig:vrot_dist_binary_binarysini}
\end{figure}

Table~\ref{tab: bayes} provides the 50th percentiles as well as the 68\%\ confidence interval of the marginalized posterior distributions of each of the parameters of Eq.~\ref{pdf_ve}. It also provides the mean ($\mu_\gamma=\alpha/\beta$) and standard deviation  ($\sigma_\gamma=\sqrt{\alpha/\beta^2}$)  of the gamma-component in the  rotation distributions. Projections of the corresponding marginalized posterior probability distributions are shown in Figs.~\ref{fig:post_binary_binarysini} and \ref{fig:post_single_randomsini}.  Finally, Fig.~\ref{fig:vrot_dist_binary_binarysini} compares an histogram of the observed \vrot\ data with the obtained best-fit rotational velocity distributions  (see Appendix~\ref{app: bayes} for details).

For the single star distribution, the best-fit parameters are different from those obtained by \citet{ramirezagudelo}. This likely originates  from the different analytical forms and  fitting methodology used. It may also have origins in the derivation of the \vrot\ measurements (now based on FWHM for consistency). Still, the overall shape of the recovered intrinsic rotational velocity distribution is in good agreement (see Fig.~\ref{fig:conv}), which allows us to validate the method. 

The Bayesian analysis allows us to confirm and quantify our previous conclusions:
\begin{enumerate}
\item[(i)] The peak of the intrinsic rotation distribution ($\varv_\mathrm{e}$) is shifted by almost 50\,\kms\ towards larger values in the sample of primaries;

\item[(ii)] The main peak is 50\%\ broader in the distribution of primaries ($\sigma_\gamma=59$\,\kms\ compared to 41~\kms\ in the single star distribution);

\item[(iii)] The high-\vrot\ tail is very limited or altogether absent in the sample of primaries\footnote{The Gaussian-component -- used to add enough degrees of freedom in the fit to be able to model a possible high-\vrot\ tail -- is not significant, at the 5\%\ significance level, in the  distribution of primaries. This yields very broad marginalized posterior distributions for $\mu$ and $\sigma$, hence poorly defined parameters for the Gaussian component as indicated in Table~\ref{tab: bayes}. Within the 68\%\ confidence interval,  the amplitude of the Gaussian component does however reach $1 - I_\gamma=0.21$, in which case one could consider it to be important. However, inspection of Fig.~\ref{fig:post_binary_binarysini} reveals that the parameters of a Gaussian that would have $1- I_\gamma \approx 0.21$ are given by $\mu \approx 200$ and $\sigma \approx 50$, i.e., such a Gaussian does not contribute significantly to a high-\vrot\ tail above $\approx 250$~\kms.}.
\end{enumerate}

Our analysis is the most complete, bias-corrected and consistent comparison of the rotation properties of single and binary massive stars. It reveals small but fundamental difference between both samples, which we will discuss in  Sect.~\ref{sec:discussion}. We also derive a suitable analytical representation of the intrinsic distribution of rotation rates that can be used, e.g., in population synthesis. In doing so, two aspects have to be kept in mind: (i) the derived distributions are present-day distributions, not initial ones; (ii) the lack of extremely slow rotators ($\vrot \lesssim 40$\kms) in the distributions results from the limitation of our \vrot\ measurement method. Higher resolution data and, for the binaries, spectral disentangling, would be needed to further our analysis at low spin rates.

\subsection{Secondaries vs. primaries} \label{subsec:secondaries} 
\begin{figure}
\centering
\includegraphics[scale=0.4]{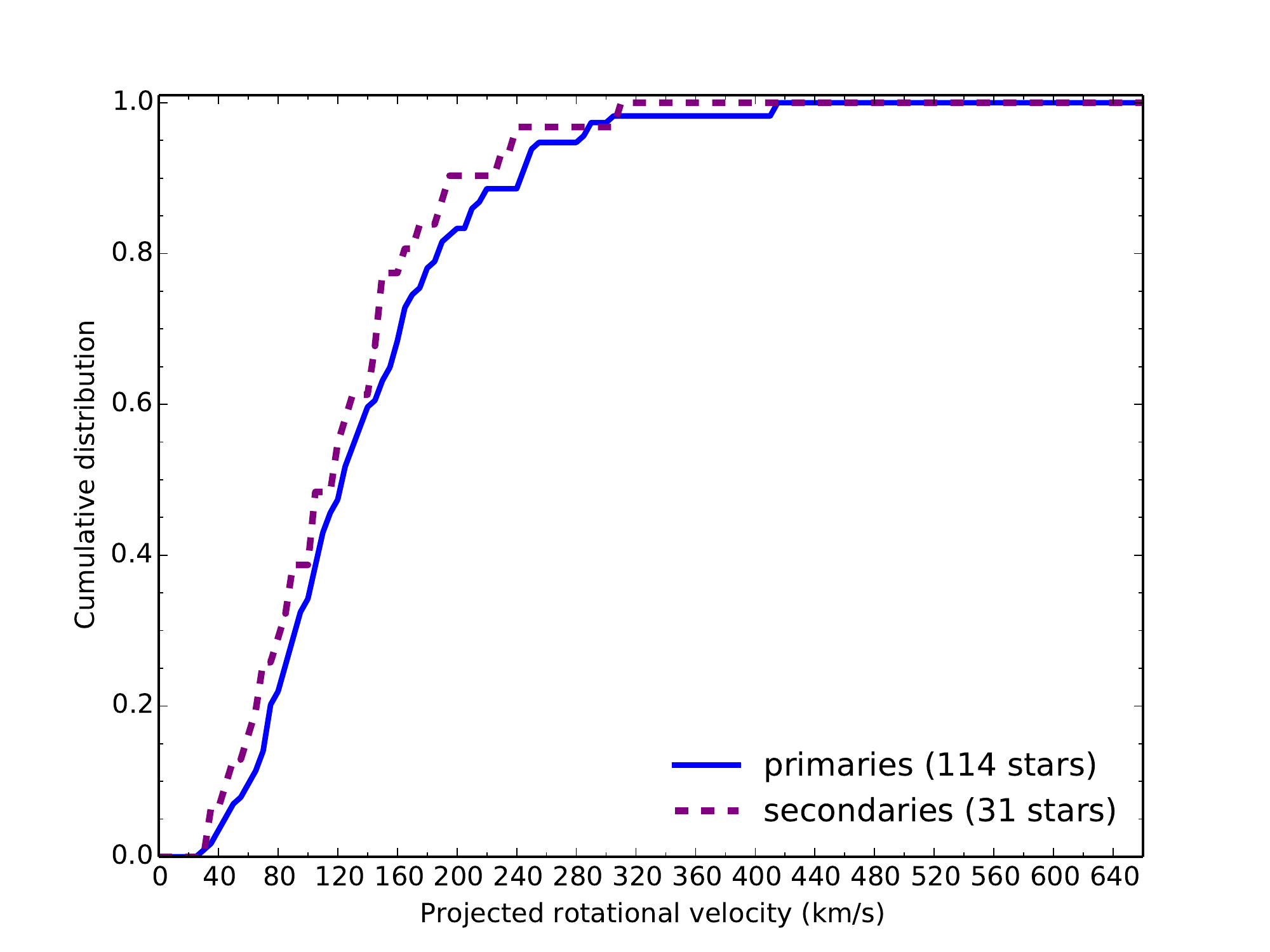}
\includegraphics[scale=0.4]{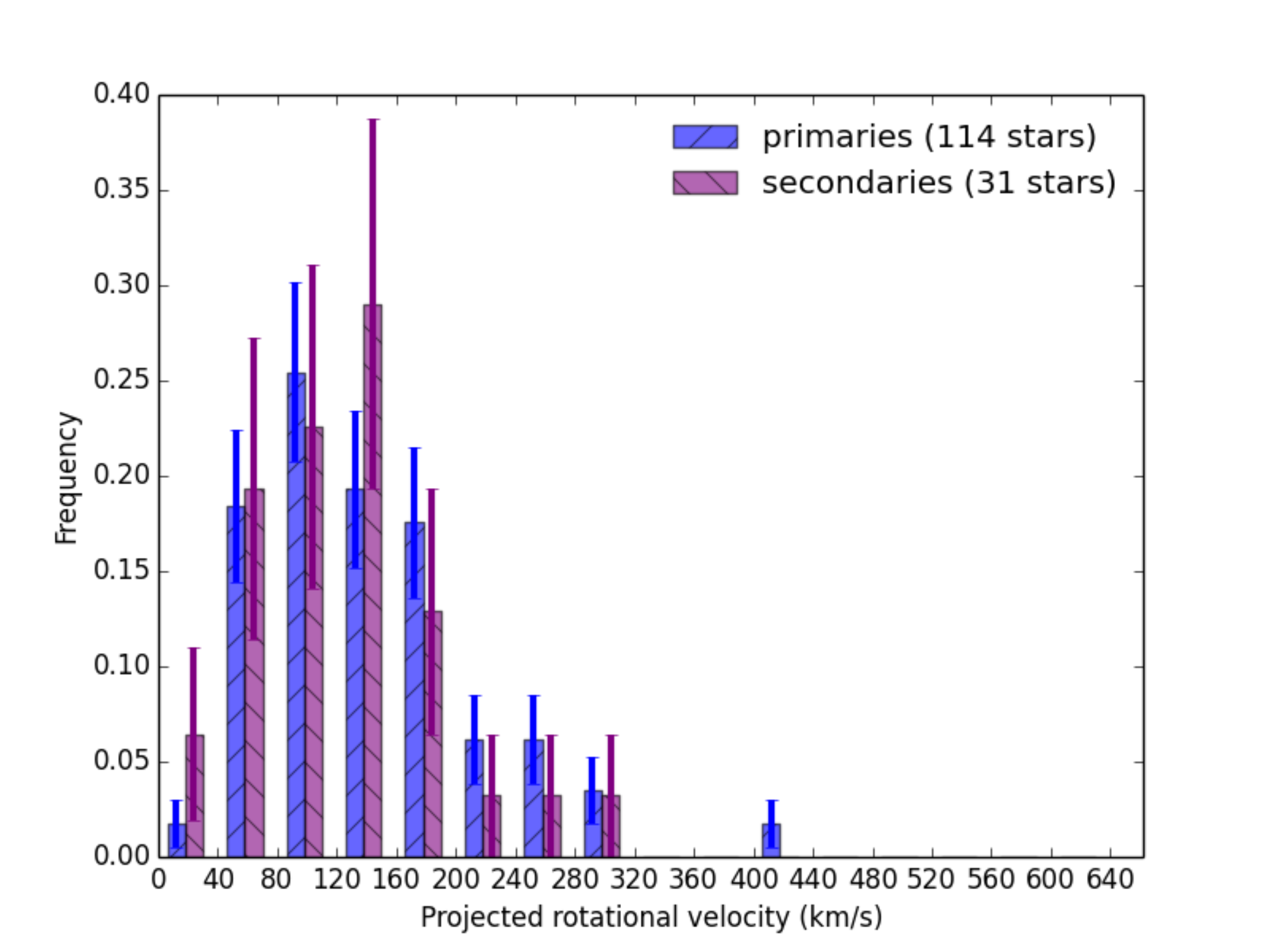}
\caption{Cumulative (upper panel) and frequency (lower panel, with Poisson error bars)
distributions of the projected rotational velocities of the O-type primaries and secondaries.
}	
\label{fig:dist_single_secondaries}
\end{figure}

In Fig.~\ref{fig:dist_single_secondaries} we show the distributions for primaries (114 stars) and secondaries (31 stars). 
Qualitatively, there is no clear difference between both distributions, i.e., the peak 
is at a similar rotational velocity and both distributions are equally broad.
The primary distribution presents  two stars with \vrot\, $\geq$ 400\,\kms.
These two outliers represent less than 2$\pm$2\,\% of the total primary sample and do not impact a comparison of the two distributions.
A KP test shows that both distributions are compatible with one another ($p_K = 95$\,\%). 

\begin{figure}
\centering
\includegraphics[scale=0.4]{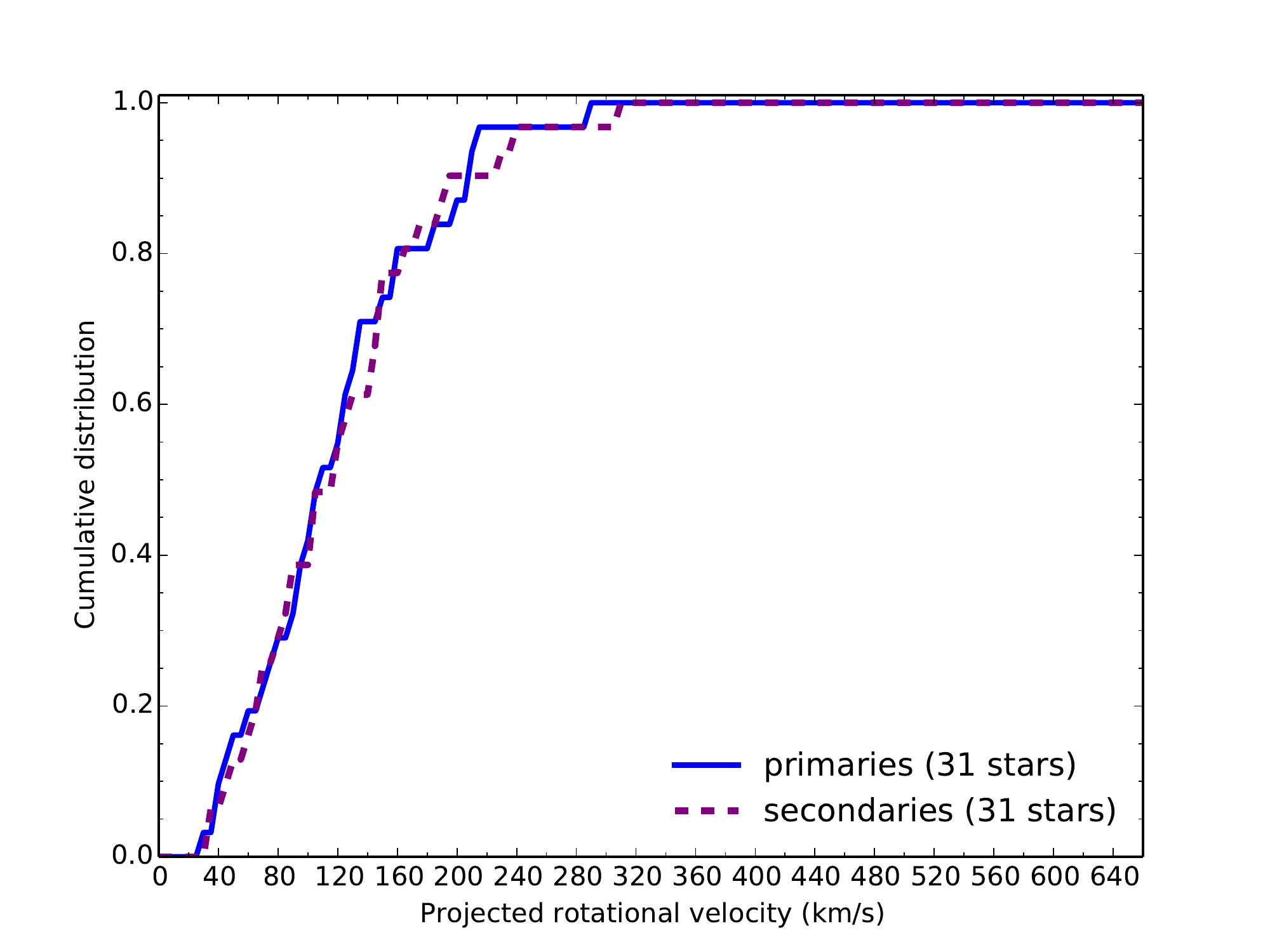}
\includegraphics[scale=0.4]{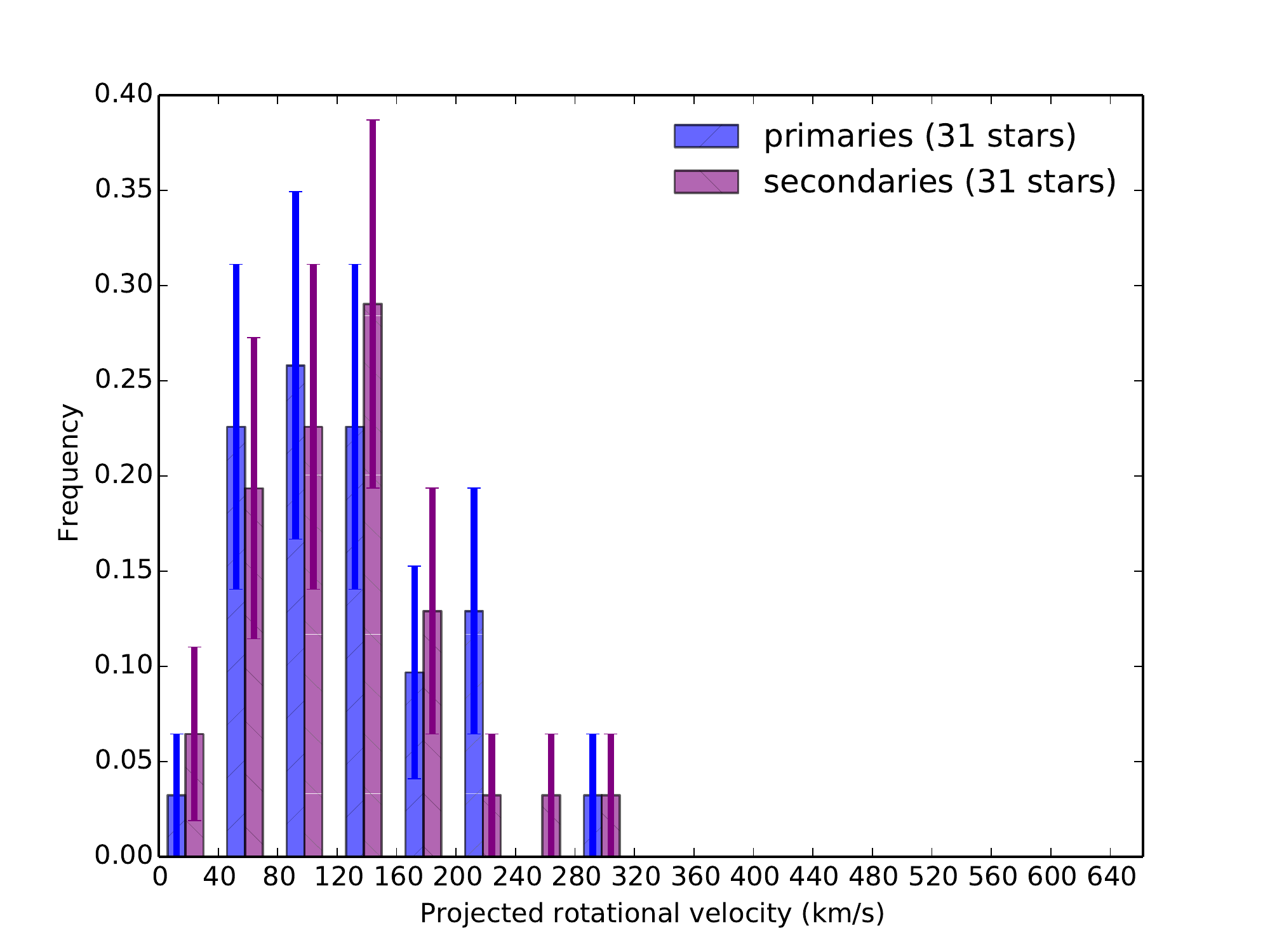}
\caption{Probability density distribution of \vrot\ for the 31 SB2 systems, for which \vrot\ of both the primaries and secondary
	       component can be constrained.
}
\label{fig:prim_sec}
\end{figure}

Similarly, Fig.~\ref{fig:prim_sec} displays the probability density distributions of the primaries and of the secondaries in our sample of 31 
SB2 binaries.  These systems mostly have large radial velocity variations ($\Delta$RV $>$ 200 \kms) indicating that the population is dominated by
close binaries. A Kuiper test yields $D$ = 0.16 and $p_{\rm K}$ = 99\%.
This is thus consistent with the hypothesis of tidal synchronization,
assuming the sizes of the accompanying primaries and secondaries are comparable.

\subsection{Small vs. large RV variation systems} 
\label{subsec:closewide} 

The size of the sample of primaries allows us to explore \vrot\ distributions for sub-populations.
By selecting a limit for the amplitude of the radial velocity variations $\rm{\Delta RV_{limit}}$ of 200\,\kms\ we divide the primary sample into two sub-samples,
i.e., stars that display $\rm{\Delta RV}\leq \rm{\Delta RV_{limit}}$ (henceforth low-$\rm{\Delta RV}$) and  stars with $\rm{\Delta RV}> \rm{\Delta RV_{limit}}$ (high-$\rm{\Delta RV}$).
The latter are systems for which  tidal synchronization is expected to be 
important during the main-sequence phase. Indeed, given our sample properties and using Monte Carlo simulations described in Appendix~\ref{appendix:B}, we can show that 88\% of systems with $\rm{\Delta RV_{limit}} > 200$\,\kms\ have orbital periods less 
than 10\,days (see section~\ref{subsec:sync_binaries}).
Most of the systems in the low-$\Delta$RV sub-sample will not suffer from tides -- 75\%\ of them are expected to have $P>10$~d and/or $q<0.5$. The  exception is of course for short-period similar-mass systems that are seen at low inclination angles or for which the RV-curve has been poorly sampled. 

\begin{figure}
\centering
\includegraphics[scale=0.4]{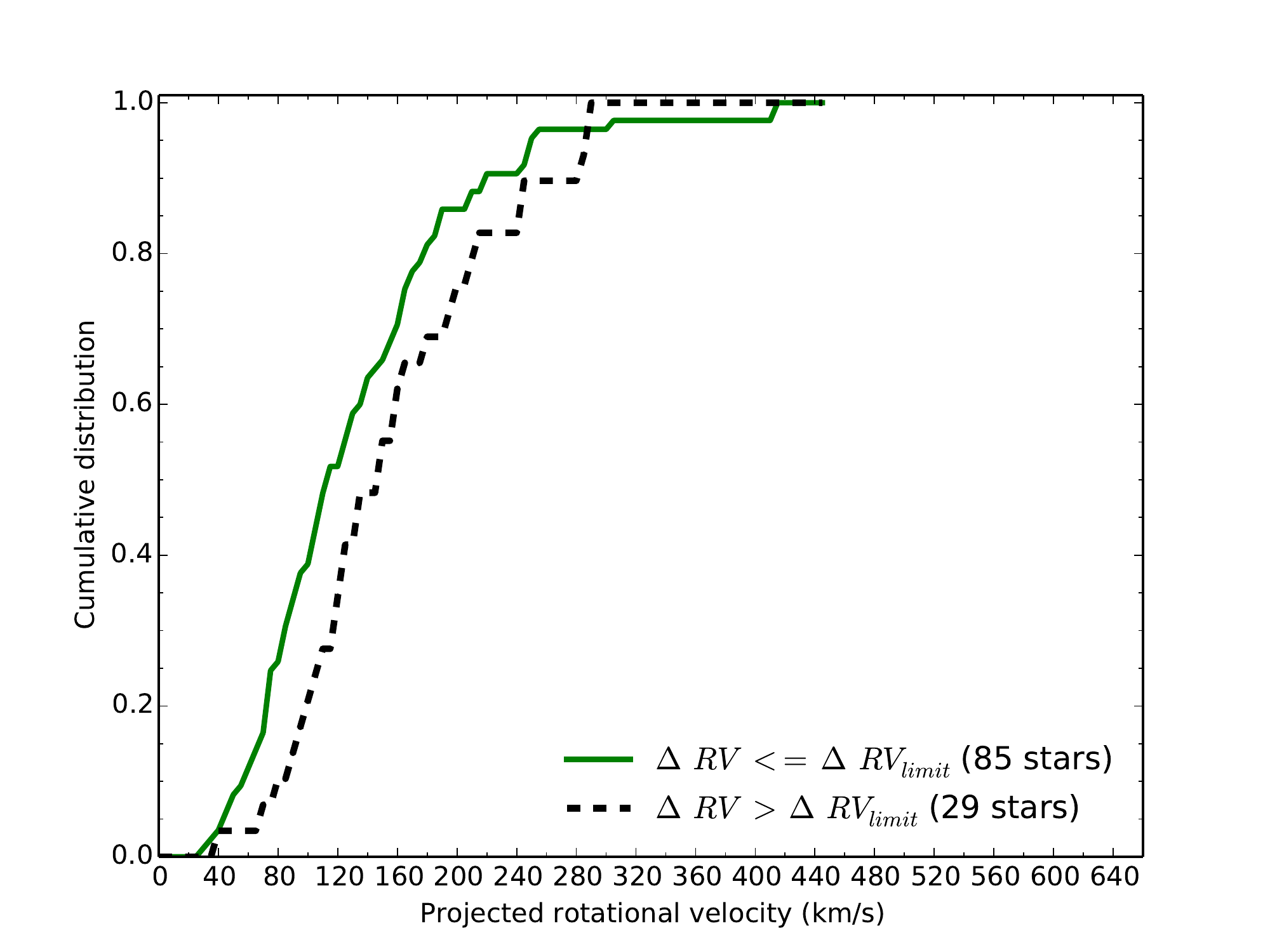}
\includegraphics[scale=0.4]{dist_binaries_new}
\caption{
Cumulative (upper panel) and frequency (lower panel, with Poisson error bars)
distributions of the projected rotational velocities of the O-type 
primaries  with  $\rm{\Delta RV}\leq \rm{\Delta RV_{limit}}$ (low-$\rm{\Delta RV}$) and   $\rm{\Delta RV}> \rm{\Delta RV_{limit}}$
(high-$\rm{\Delta RV}$). 
}
\label{fig:dist_binaries}
\end{figure}

Figure~\ref{fig:dist_binaries} plots the \vrot\ distributions of the low-$\rm{\Delta RV}$ (85 stars) and high-$\rm{\Delta RV}$ (29 stars)
sub-samples. 
At low rotation velocities  the distribution of high-$\rm{\Delta RV}$,
compared to the low-$\rm{\Delta RV}$, appears shifted by one bin ($\approx 40$~\kms) to higher rotational velocities. 
The high-velocity tail (i.e., \vrot\, $>$ 300 \kms) of the low-$\rm{\Delta RV}$ distribution has three primaries (3.5$\pm$2.4\,\% of that sample), while   there is no star in the high-$\rm{\Delta RV}$ distribution.
Despite these apparent 
differences, a KP test does not allow us to reject the hypothesis that both samples are drawn from the same parent distribution ($p_K = 57$\,\%),
possibly as a result of the limited size of the high-$\rm{\Delta RV}$ sample. The weighted mean of the samples (130$\pm$2~\kms\ and 153$\pm$5\,\kms\ 
for the low- and high-$\rm{\Delta RV}$ samples, respectively) are however significantly different from one another, confirming 
a higher average spin rate of
the high-$\rm{\Delta RV}$ (short periods) systems.

\subsection{Single stars versus  low-$\rm{\Delta RV}$ binaries} \label{subsec:singlelowrv} 

\begin{figure}
\centering
\includegraphics[scale=0.4]{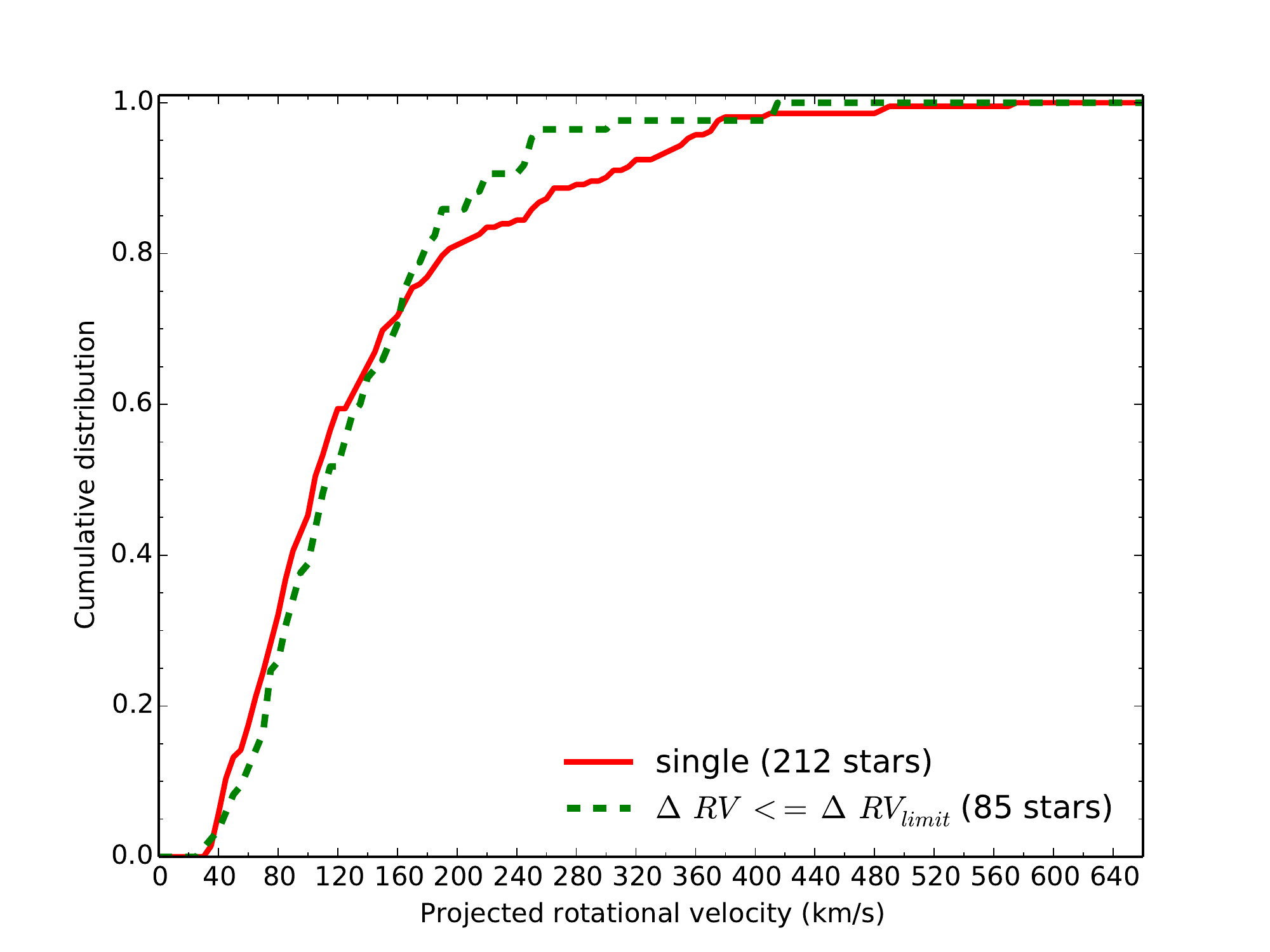}
\includegraphics[scale=0.4]{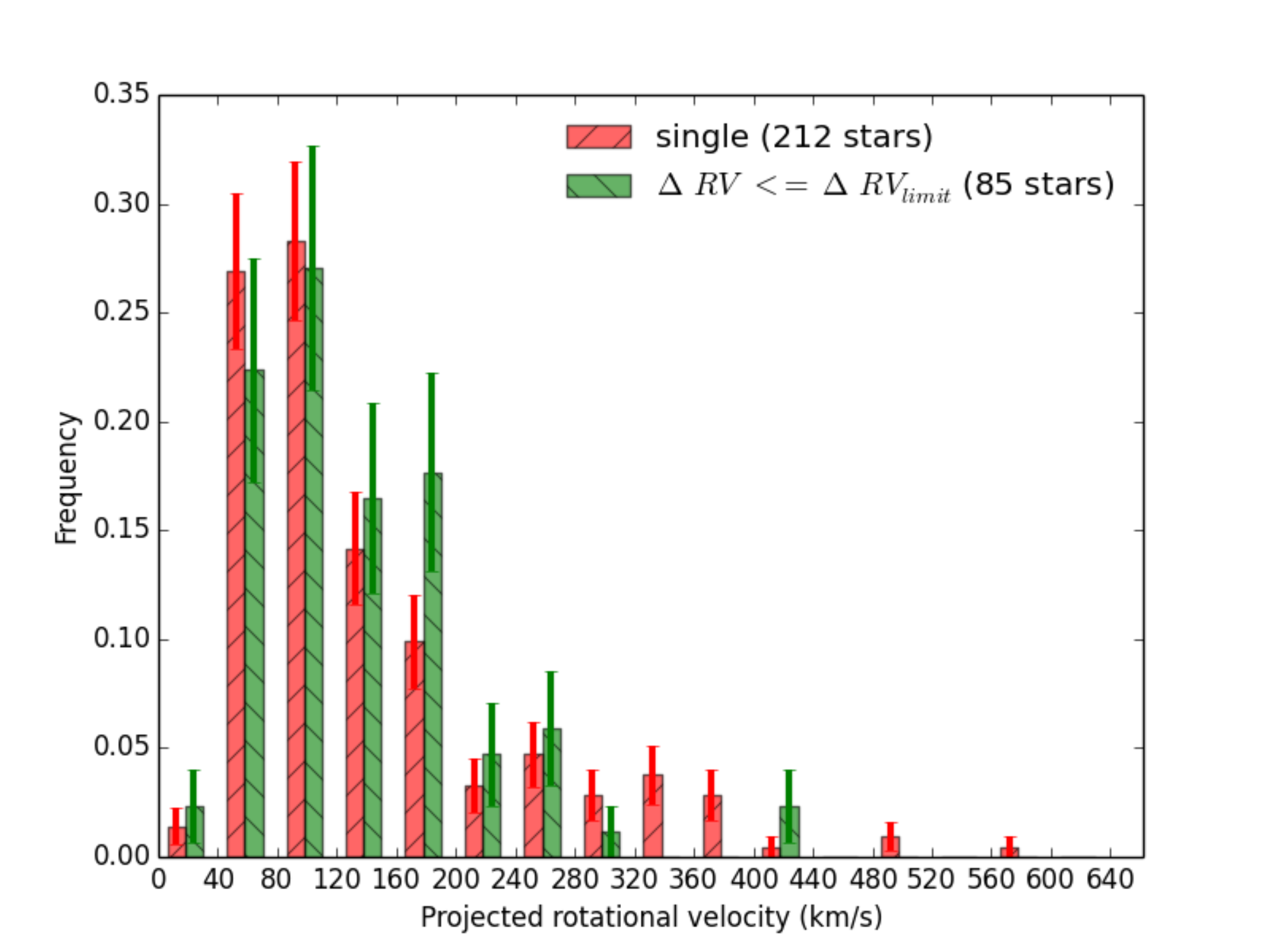}
\caption{
Cumulative (upper panel) and frequency (lower panel, with Poisson error bars)
distributions of the projected rotational velocities of the presumed-single O-type stars 
and the sub-sample of low-$\rm{\Delta RV}$ sources.
}
\label{fig:dist_single_rvlow}
\end{figure}

We compare the sub-sample of low-$\rm{\Delta RV}$ sources with that of the presumed-single stars in Fig.~\ref{fig:dist_single_rvlow}.
As pointed out in Sect.~\ref{subsec:closewide} the primary stars in the sample of long period systems are not likely to interact with their
companion during their main-sequence lives (where we have used a 
low-$\rm{\Delta RV}$ as a proxy for long periods, barring low inclination systems and orbit sampling effects; see Sect.~\ref{subsec:closewide}).  As the bulk of the binary systems and the
presumed-single systems are on the main sequence, a comparison of the two  may thus help to address the question
whether or not the primaries in wide (enough) binaries and presumed-single stars leave the formation process with similar spin-rate
distributions.  
In \citet{ramirezagudelo} we claimed the low-velocity peak to be the representative of the outcome of the formation process
where in (projected) rotational velocity it extended up to $\sim$\,160\,\kms.\,
Indeed, for $\vrot \leq 170$\,\kms\ both distributions are very similar and a KP shows 
that both distributions are compatible with each other ($p_K = 76$\,\%, see Table~\ref{table:2}).

For the binaries with projected rotational velocities larger than 170\,\kms\ both distributions behave differently. The sample 
of presumed-single stars presents a significant  contribution of stars rotating more rapidly than 300\,\kms, 
while the wide-binary sample does not. 

 \section{Discussion}\label{sec:discussion}
  
The dominant feature in the \vrot\ distributions of the O-type primaries and secondaries
in 30 Dor is their  low-velocity peak. Compared to the rotational distribution of single stars,  there is a low occurrence of stars spinning faster than 300 \kms\ in the distribution of primaries. The well developed high-velocity tail seen in the presumed-single stars is thus nearly absent.
In this section we consider possible physical mechanisms that may be impacting the global shape of the initial
\vrot\ distribution of binary components,
notably tidal interaction and mass transfer.  

\subsection{Signatures of tidal interaction}
\label{subsec:sync_binaries}

As discussed earlier, the bulk of our binary sample is expected to consist of pre-interaction systems. 
\citet{selma} study the effects of binary evolution on the rotation rates of primaries and secondaries as a function of
initial period and initial mass ratio. 
Prior to the primary star filling its Roche lobe, changes in stellar structure,
mass loss and tides may affect the spin rates. 
%

Following \citet{weiderandvink2010}, we estimate that about 20\% of the stars in our total sample have masses
above 40\,\msun, with the most massive stars (accounting for $\sim$4\% of the total) being roughly  60--80 \,\msun.  It is
these stars that develop winds that are strong enough to facilitate significant angular momentum loss during their
main-sequence evolution \citep[see][]{2001A&A...369..574V}.
%
As we focus on the ensemble properties, this implies that the bulk of the stars will not have stellar winds strong
enough to spin them down significantly on the main-sequence timescale.
Exceptions may be stars with strong large-scale magnetic fields,  which we discuss separately below.

  Stellar expansion due to changes in internal
structure will only have a minor effect on the spin rate when the star approaches the end of its main-sequence lifetime,
and can be further neglected \citep{brott,selma}.  The most important effect on the spin rate before Roche lobe overflow 
takes over is therefore that of tides. 
In systems where the
radii of the components is comparable to the separation between the stars \citep{selma}, tidal interaction 
strives to synchronize the rotation of the binary components with the orbit. For massive binaries this
corresponds to systems with orbital periods up to about a week.  Once synchronized the rotational velocity of
the co-rotating primary increases 
mainly because of secular stellar expansion. 

Like the single O-type stars in 30~Dor, the \vrot\ distributions of primaries and secondaries also present a 
low-velocity peak at around 100\,\kms, though it is broader and skewed to somewhat higher spin rates
(see Figs.~\ref{fig:dist_single_primaries}-\ref{fig:prim_sec}). Under the assumption that
the initial spin distributions of binary components and single stars are the same, this difference is consistent with tidal interaction
playing a role in our binary sample.  
Effects of tides and, in a minority of cases, possible slow Case A mass transfer
\citep[see Figures 2 and 3;][]{selma}, appear 
most visible in the contrast between the projected spin-rate distributions of low-$\rm{\Delta RV}$ versus
high-$\rm{\Delta RV}$ systems (see Fig.~\ref{fig:dist_binaries}), the latter corresponding to closer binaries and
displaying larger spin rates (see Sect.~\ref{subsec:closewide}).

\begin{figure}
\centering
\includegraphics[scale=0.45]{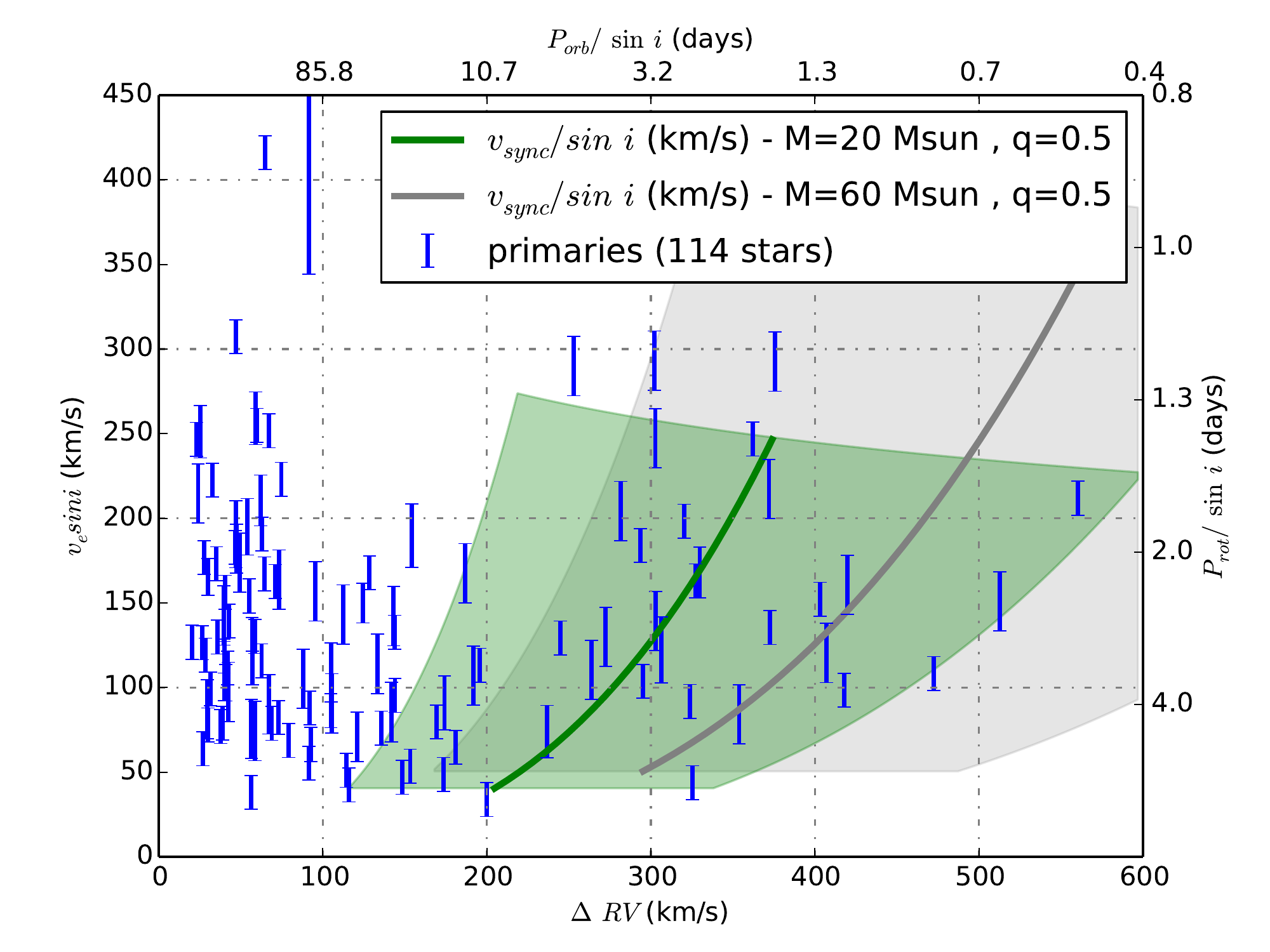}
\caption{Projected rotational velocity vs.\ amplitude of the radial velocity variations in the primary stars sample (114 stars). The shaded regions show approximately where we expect tidal synchronization to play a role during the main sequence evolution for a primary mass of 20 and 60\,\msun\ \textbf{(with $R$\,=\,8 and 10\, $R_{\odot}$, respectively)}. Here we varied the mass ratios $q$ from 0.25 and 1 and periods $P_{\rm orb}$ from 0.25 to 10 days. The labels on the top and right hand axes show the corresponding orbital and spin period for a typical circular binary system assuming a 20\,\msun\, primary and a mass ratio of 0.5.}
\label{fig:vsini_deltarv}
\end{figure}

\begin{figure}
\centering
\includegraphics[scale=0.45]{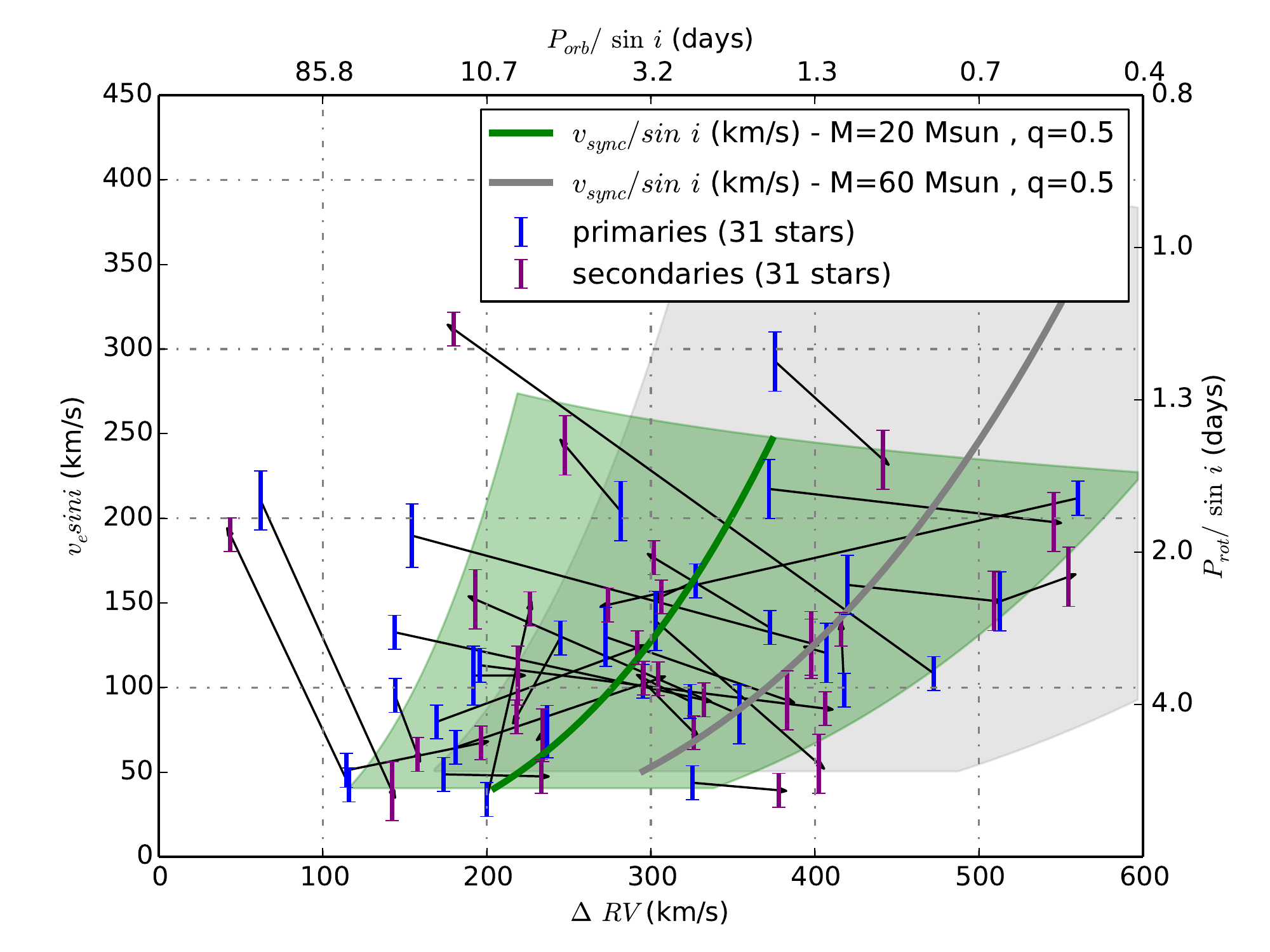}
\caption{Projected rotational velocity vs.  amplitude of the radial velocity variations for secondaries in SB2 systems (31 stars) together with their corresponding primaries, linked
with arrows. 
The definition of the green and gray regions and the labels on the top and right hand axes are as in Fig.~\ref{fig:vsini_deltarv}. 
}
\label{fig:vsini_deltarv_SB}
\end{figure}

To investigate the effect of tidal locking we make use of a diagram showing the relation between projected rotational velocity 
and amplitude of the RV variations (see Figs.~\ref{fig:vsini_deltarv} and~\ref{fig:vsini_deltarv_SB}). The timescale of 
tidal synchronization is a function of the primary mass $M_{1}$, the mass ratio of the primary and secondary $q = M_{2}/M_{1}$, and
the orbital period $P_{\rm orb}$.  For our sample we lack information about the latter and therefore we use the amplitude of the radial velocity variations $\Delta$RV as a proxy of the period.   

For a (hypothetical) fully synchronized binary system  \vrot\ and $\Delta$RV are fully determined by  $P_{\rm orb}$, $M_{1}$, the radius $R$ of the primary and $q$ through Kepler's law. The thick green line in Fig.~\ref{fig:vsini_deltarv} indicates this relation for a typical system assuming  a synchronized orbital and spin period between 10 days (lower left) down to the minimum period at which the primary star would be filling its Roche lobe (upper right).  Here we adopted a primary mass of 20 \msun, with $R$\,=\,8\, $R_{\odot}$, and a mass ratio $q=0.5$.  The shaded region shows the area spanned by similar relations varying the mass ratio between $q=0.25$ (left side boundary of the green area) and $q=1$ (right side boundary).   The thick gray line and grey shaded area show the same adopting a primary mass of 60 \msun\, with $R$\,=\,10\, $R_{\odot}$.  It stretches out to both higher rotation rates, because \vrot\ scales with the stellar radius which is larger for more massive stars and to larger $\Delta RV$ which is proportional to  $M_{1}^{1/3}$.

Tides are most effective for short periods systems  \citep{Zahn75, Zahn77} with a weak preference for equal mass systems. Stars in the upper part of the shaded areas are expected to be fully synchronized as the timescale for synchronization becomes (much) less than 1 percent of the main-sequence lifetime \citep{selma2009b}. For binary systems with periods larger than about 10 days and/or extreme mass ratios we do not expect tides to have a significant effect during the main sequence evolution.

Forty VFTS primaries fall within the shaded zones.  According to the above argument, they are expected to be synchronized or to tend
toward synchronization. Given their \vrot\ properties, these systems are sufficiently numerous to explain the overpopulation of primaries in the 100-300~\kms\ range compared to the single star sample (Sect.~\ref{subsec:primaries}).




For the SB2 systems one can in principle check whether the two stars have synchronized. In Fig.~\ref{fig:vsini_deltarv_SB}
we show the 31 stars for which we have obtained $\Delta RV$ and \vrot\ for both the primary  and the secondary components.
Most of the systems fall in the green zone. This is a selection effect as SB2s are most easily recognized if the
radial velocity shifts are large \citep{sana2011}.  All primary and secondary pairs have been connected with a straight line. 
If a pair is synchronized the angular spin velocities are the same. In that case one expects the lines to be horizontal in
case of equal sized stars, and about horizontal when allowing for size differences. Indeed, by far the most systems 
show an about horizontal line connecting the two components
and we thus interpret this property as the signature of tidal synchronization. Two distinct outliers are both located
in the lower left corner of the diagram, where the timescale for tidal synchronization is the longest.

Using population synthesis \citet{selma} show that tides are expected to only lead to
relatively few systems 
with velocities larger than 200 \kms, and that $\sim$300 \kms\ is the limit that tides can achieve.  Close binary systems that 
potentially could have acquired spin rates much in excess of 300 \kms\ at birth would be spun down to $\sim$200
\kms. Indeed, this expectation
too seems consistent with our findings: none of the components of the close systems (using $\Delta$RV $>$ 200 \kms\ as a
proxy for identifying such systems) spin faster than this limit (see Sect.~\ref{subsec:tail}).
Only when mass transfer occurs, later-on in the evolution, the secondary may reach spin rates of up to about 600 \kms.

\subsection{Magnetic braking}
\label{subsec:magnetic}

If a massive star possesses a sufficiently strong magnetic field, the interplay
of the stellar winds with the stellar magnetosphere may lead to a rapid spin-down of the star.
The efficiency of the spin-down depends both on the strength of the magnetic field and on that of the winds \citep{2009MNRAS.392.1022U}. For known magnetic galactic O stars, the timescale of magnetic spin-down  is of the order of several 0.1~Myr to 3~Myr \citep{2013MNRAS.429..398P}. Given the lower metallicity, hence lower  wind strengths of stars in the 30~Doradus region, their spin-down timescales will be longer. 
Using \citeauthor{2009MNRAS.392.1022U} formalism, we estimated them to be of the order of 6 to 25~Myr for a 1~kG field, and a factor of ten shorter for a 10~kG field.

Possibly up to 10\% of the Galactic O stars host a long-lived large scale magnetic field \citep{2014arXiv1408.2100M,2014IAUS..302..265W}. However, only one has a known magnetic field stronger than 10~kG \citep[NGC\,1624-2,][]{2012MNRAS.425.1278W, 2013MNRAS.429..398P}, a condition that seems required but rarely met for magnetic braking to efficiently affect LMC O stars. While the detected magnetic fields in Galactic OB stars are believed to be in a stable configuration,  Fossati et al.\ (2015, in prep.) argue that the decay timescales  in massive main-sequence stars may be shorter than their lifetimes. If the field decay timescale is longer than the magnetic breaking one (for stars to have time to spin-down), but fast enough for most strong magnetic fields to elude detection,  a substantially larger fraction of stars may have been affected by magnetic breaking than what the current detection rate of strong magnetic fields suggests. As a consequence one needs to consider the possibility that magnetic braking may have some impact in our sample. 

The origin of the magnetic fields in massive stars is still under debate. One possibility is that of a fossil field that is frozen into a stable large-scale configuration during the star-formation process
\citep{2010A&A...517A..58D}. Another 
 is that the fields are formed
during a strong binary interaction, e.g., a stellar merger \citep[e.g.,][]{2014MNRAS.437..675W}.
Merger products are not expected to be found in close binaries,
as then the pre-merger triple star would have been highly unstable. 
This agrees with the fact that the vast majority of magnetic massive stars
seems to lack a close binary component, in contrast with their non-magnetic
counterparts \citep{2014arXiv1409.1094A}.

If magnetic spin-down is affecting our sample and if magnetic fields are generated by binary interaction, one thus expects a reduced number of slowly spinning components in close binaries compared to single stars or to wide binaries.  However, tidal interaction in close binaries  leads 
to synchronization, such that this effect is not detectable.
The resulting slowly spinning post-interaction products would mostly 
show up in the sample of presumed-single O stars. This would move
the main peak in their initial spin distribution towards lower spin velocity. 
A similar effect may affect the wide binary distribution. Indeed, systems born as hierarchical triples  may now appear as a wide binary, 
after the close pair has interacted. The post-interaction products may spin slowly because of magnetic spin-down.

To conclude, this discussion illustrates the possibility that the observed shift of the spin distribution main peaks in  the
primaries and in the single stars samples may partly be
due to magnetic braking in post-binary interaction products.
Similarly  magnetic braking may also have contributed to the larger average spin rate observed in  the short- compared to the long-period systems. To test this scenario, further higher resolution spectra would be necessary as they would enable the search for  magnetic spin-down signatures in the form of sub-structures in the low-velocity part of the spin distributions.

\subsection{The nature of the high-velocity tail and the single-star channel toward long-duration GRBs}
\label{subsec:tail}

\citet{ramirezagudelo} suggested that the high-\vrot\ tail identified in the projected spin-rate distribution of the presumed-single O-type stars in the VFTS is populated by binary interaction products that have been spun up through mass transfer or coalescence. A test of this scenario is to determine the projected spin-rate distribution of primaries and to check whether a high-velocity tail is
present or not.  If the tail in the sample of presumed-single stars is dominated by post-interaction products one may anticipate that it is less pronounced or even absent
in the binary sample. Indeed tides in pre-interaction binaries can not spin the components up beyond 300 \kms\ (see 
Sect.~\ref{subsec:sync_binaries}).  We find that 2.6$\pm$1.8\,\%  of the primaries spin faster
than 300 \kms, compared to 10.4$\pm$1.9\,\%  in the presumed-single sample. 
This supports the hypothesis that the high-velocity tail
is actually dominated by post-interaction products and that genuine single stars spinning faster than 300 \kms\ are rare
and maybe absent altogether.  

Rapidly spinning single stars have been considered as a potential channel towards the production of long-duration 
gamma-ray bursts (GRBs; \citeauthor{yoon2005} \citeyear{yoon2005}, \citeauthor{woosley} \citeyear{woosley}).  If stars are born 
with rotational speeds of at least 300--400 \kms, depending on metallicity, they will evolve quasi-chemically homogeneously 
\citep[e.g.,][]{1987A&A...178..159M,brott}, i.e., they remain compact throughout their hydrogen burning phase and become
Wolf-Rayet stars directly afterwards.  In this way their stellar cores may retain enough angular momentum to fulfill the requirements of the 
collapsar model towards GRBs \citep{1993ApJ...405..273W}, especially in low-metallicity environments where stellar winds of 
Wolf-Rayet stars are weaker, hence remove less angular momentum \citep{vink2005}.  

If rapidly spinning single stars are rarely formed 
or not formed at all this particular channel toward GRB progenitors may not occur in nature.  It is however too early to tell whether 
or not this implies that single stars may not produce GRBs altogether, as stars with low magnetic torques \citep{2011arXiv1105.4193W} 
or core collapse early-on in the Wolf-Rayet phase \citep{vink2011a,grafener2012} might provide alternative routes. 
Spun-up mass gainers in close binaries might evolve chemically homogeneously and provide
a GRB at the end of their life \citep{2007A&A...465L..29C}, such that the total
number of GRBs predicted from chemically homogeneous evolution does not necessarily change.

\subsection{The nature of the low-velocity peak and the formation of massive stars}
\label{subsec:formation}

The low-$\Delta$RV sources are mostly SB1 systems with 
relatively long orbital periods
(with probable periods $\sim$10-1000 d) and/or low mass ratio systems, neglecting inclination
and sampling effects. They are likely dominated by pre-interaction systems that do not undergo significant tidal synchronization (Fig.~\ref{fig:vsini_deltarv}). The few systems that are located in the region of the parameters where tidal interaction is expected to take place (shaded areas in Fig.~\ref{fig:vsini_deltarv}) occupy the lower left corner, where the tidal synchronization timescale is the longest.  Moreover, the stellar winds of the bulk of this sample are too weak to induce significant angular-momentum 
loss, and spin-down due to envelope expansion as a result of secular evolution may be neglected.  This points to the idea
that the spin rates of the low-$\Delta RV$, wide and low-$q$ spectroscopic binaries have seldom been modified since the stars were formed and thus reflect the outcome of the formation process.

While the formation process of massive stars is heavily debated 
\citep[e.g.,][]{zinnecker2007,TBC14,2014arXiv1403.3417K}, most theories agree on the need for disk-mediated accretion.  In its simplest form, the collapse of the initial natal cloud and the gas-accretion phase 
concentrate by far more angular momentum in the central region than can be stored in a single star. Therefore,
the spin rate of genuinely single massive stars may naively be expected to be close to the star break up velocity (the so-called angular momentum problem).  
Gravitational torques between star and disk appear capable of limiting the stellar spin rates to about half the break-up speed $\varv_{\rm crit}$
\citep{2011MNRAS.416..580L}. This value is however still substantially higher than the 0.1--0.3 $\varv_{\rm crit}$ that is typical for the observed low-velocity peak 
\citep[e.g.,][]{conti1977,huanggies1,penny2009,ramirezagudelo}.  Magnetic breaking might aid in a further spin-down of the stars, either by self-induced fields or by fossil fields. 
Alternatively, the formation of massive stars as part of a multiple system may provide a 
reservoir to store angular momentum from the natal cloud into the orbit of the components. 
This too may alleviate the need to have a large initial stellar spin.

The low-velocity peak of the presumed-single stars studied by \citet{ramirezagudelo} 
may also reflect
the initial conditions. 
Here we compare the low-\vrot\ regime of the two samples to assess whether the spin distribution of presumed-single stars is similar to that of low-$\Delta$RV primaries in
binary systems.  A Kuiper's test selecting the stars having $\vrot \leq 170$ \,\kms\ in both samples fails to identify significant differences 
and we henceforth accept that these samples may be drawn from the same parent distribution. 

The resemblance of the  low-velocity peak in the \vrot\ distributions of the long-period and/or low-$q$ spectroscopic binaries and the presumed-single 
stars thus suggests similarities in their formation process. This could be the case if primordial
 magnetic fields control the birth spin properties of massive stars, be they either single or part of a binary system.  
If the initial spin properties are controlled by multiplicity, i.e., by a formation process that proceeds
through the creation of pairs or higher order multiples, the similarity in the spin distributions of presumed-single stars and primaries
might suggest that true single stars do not actually exist.
Interestingly, a high angular resolution survey of galactic massive stars combined with spectroscopic results from the literature  has revealed 
an extreme multiplicity fraction ($>$90\%), suggesting that  all massive stars may form as part of a binary or higher order system and that higher hierarchical systems are not uncommon \citep{2014ApJS..215...15S}. 
Therefore, genuine single massive stars may not exist.

\section{Conclusions}\label{sec:conclusions}

Based on a sample of presumed-single stars in the 30 Doradus region in the LMC presented by  \citet{ramirezagudelo} we have constructed 
a calibration that allows to estimate the projected spin rates of O-type stars from FWHM measurements.  We have applied this 
$\vrot$--FWHM calibration to determine the projected spin rates
of the components of 83 SB1 and 31 SB2 O-type binary systems in 30 Dor.

Given the uncertainties in individual measurements (20--40~\kms)
we have focused on the spin-rate distribution, and have compared it to that of presumed-single stars. 
The sample is large enough to define sub-samples that could be compared to each other.
This has lead to the following conclusions:  

\begin{enumerate}
\item The overall spin distribution of the  O-type binary components qualitatively resembles
	 that of the single O stars in the 30~Dor region \citep[see][]{ramirezagudelo}. Both are dominated by a low-velocity peak 
	($<$ 200~\kms) which contains $\sim$80\% of the samples, and both show a shoulder at intermediate velocities 
	($200~\kms \leq \vrot \leq 300$~\kms). However, the spin distribution of presumed-single stars features a
	clearly identifiable high-velocity  tail ($>$ 300~\kms), which is hardly populated in the binary population.\\


\item The main peak of the binary distribution, centered at $\sim$100 \kms, is broader and skewed to somewhat higher spin rates
      than in the single star distribution.
          Similarly, the spin distribution of primaries in systems 
          with $\Delta \mathrm{RV} > 200$~\kms, which we associate with close binaries with $P_{\rm orb} \lesssim 10$ days,
          is skewed towards higher spin rates than that of primaries in the presumably wider systems with
          lower $\Delta$RVs.\\

\item About one third of our binaries are so close that the spins of their components are expected to be tidally synchronized with the orbital motion. The current spin rates in such close binaries are thus expected to be independent of their birth spins. Tidal synchronization may be responsible for the  larger average  spin rates that we observe between components of close binaries and that of single stars and of stars in longer period systems. A part of this shift might also result from magnetic braking affecting a fraction of the single-star and long-period binary population. The latter scenario would be consistent with magnetic fields being generated by strong binary interaction in an earlier epoch of evolution of these systems.\\

\item The SB2 systems in our sample mostly show radial velocity variation amplitudes above 200 \kms\ and have tidal synchronization timescales
          that are shorter than their main-sequence timescale.  On average, we find that 
          the \vrot\ values of SB2 primaries and secondaries are
          similar, as expected for tidally synchronized binaries.\\

\item  Among the primaries of our binary systems, stars spinning faster than $\vrot = 300$\,\kms are about
       four times as rare as in the single star population, in relative terms.
          This is consistent with the hypothesis, proposed by 
	 \citet{ramirezagudelo} and \citet{selma}, that this tail in the (presumed) single star distribution is dominated by
	 post-interaction binary products. 
	 In particular, the population of wide binaries shows hardly any components spinning faster than 300 \kms\, suggesting that
	 massive star formation rarely produces such systems, if at all.\\

\item The low-velocity peak in the sample of primaries in wide spectroscopic binaries, that orbit their companion at distances
	of the order of 1--10 AU, is very similar to that of the presumed-single stars.  
        This, together with the overall qualitative similarity in the spin distribution of single stars and binary components,
        and plausible scenarios for even explaining the quantitative differences through binary evolution,
        suggests that the spins of massive stars are set independently of whether they form in 
        single stars or binary systems and are controlled by similar physics.

\end{enumerate}

The results presented in this work focus on the overall spin distribution of our binary O-star sample. A detailed analysis of the 
properties of multiple systems -- yielding periods rather than $\Delta$~RV  as proxies of these periods -- and their relation to
stellar parameters is a logical next step in assessing the role of tides and mass-transfer in binary systems and the differences and similarities in the  intrinsic rotational properties of single and binary systems. Higher spectral resolution data and an emphasis on metallic lines would also offer the possibility to obtain more accurate measurements in the low spin regime. These are needed to investigate possible sub-structures in the low-velocity peak 
and their relation to, e.g.,  magnetic spin-down.

\begin{acknowledgements}
The authors thank Sergio Sim\'{o}n-D\'iaz for constructive discussions.
S.d.M. acknowledges support by NASA through an Einstein Fellowship grant.
VHB is grateful to the ``Fonds the recherche du Qu\'ebec - Nature et technologies" (FRQNT) 
for financial support through a postdoctoral fellowship grant. The authors are grateful to the anonymous referee for comments and suggestions that help improving the quality of the paper.
\end{acknowledgements}




\bibliographystyle{aa}	

\begin{thebibliography}{63}
\expandafter\ifx\csname natexlab\endcsname\relax\def\natexlab#1{#1}\fi

\bibitem[{{Abt} {et~al.}(2002){Abt}, {Levato}, \& {Grosso}}]{abt}
{Abt}, H.~A., {Levato}, H., \& {Grosso}, M. 2002, \apj, 573, 359

\bibitem[{{Alecian} {et~al.}(2014){Alecian}, {Neiner}, {Wade}, {Mathis},
  {Bohlender}, {C{\'e}bron}, {Folsom}, {Grunhut}, {Le Bouquin}, {Petit},
  {Sana}, {Tkachenko}, {ud-Doula}, \& {the BinaMIcS
  collaboration}}]{2014arXiv1409.1094A}
{Alecian}, E., {Neiner}, C., {Wade}, G.~A., {et~al.} 2014, in ``New Windows on
  Massive Stars", IAU Symposium 307, Eds. G. Meynet, C. Georgy, J. Groh \& P.
  Stee, Cambridge University Press, in press (arXiv 1409.1094)

\bibitem[{{Brott} {et~al.}(2011){Brott}, {de Mink}, {Cantiello}, {Langer}, {de
  Koter}, {Evans}, {Hunter}, {Trundle}, \& {Vink}}]{brott}
{Brott}, I., {de Mink}, S.~E., {Cantiello}, M., {et~al.} 2011, \aap, 530, A115

\bibitem[{{Cantiello} {et~al.}(2007){Cantiello}, {Yoon}, {Langer}, \&
  {Livio}}]{2007A&A...465L..29C}
{Cantiello}, M., {Yoon}, S.-C., {Langer}, N., \& {Livio}, M. 2007, \aap, 465,
  L29

\bibitem[{{Carroll}(1933)}]{carroll}
{Carroll}, J.~A. 1933, \mnras, 93, 478

\bibitem[{{Conti} \& {Ebbets}(1977)}]{conti1977}
{Conti}, P.~S. \& {Ebbets}, D. 1977, \apj, 213, 438

\bibitem[{{de Koter} {et~al.}(2013){de Koter}, {Sana}, {Evans},
  {Besthenlehner}, \& {Taylor}}]{2013ASPC..470..111D}
{de Koter}, A., {Sana}, H., {Evans}, C., {Besthenlehner}, J.~M., \& {Taylor},
  W.~D. 2013, in Astronomical Society of the Pacific Conference Series, Vol.
  470, 370 Years of Astronomy in Utrecht, ed. G.~{Pugliese}, A.~{de Koter}, \&
  M.~{Wijburg}, 111

\bibitem[{{de Koter} {et~al.}(2011){de Koter}, {Sana}, {Evans}, {Bagnoli},
  {Bastian}, {Bestenlehner}, {Bonanos}, {Bressert}, {Brott}, {Cantiello},
  {Carraro}, {Clark}, {Crowther}, {de Mink}, {Doran}, {Dufton}, {Dunstall},
  {Garcia}, {Gr{\"a}fener}, {H{\'e}nault-Brunet}, {Herrero}, {Howarth},
  {Izzard}, {K{\"o}hler}, {Langer}, {Lennon}, {Ma{\'{\i}}z Apell{\'a}niz},
  {Markova}, {Najarro}, {Puls}, {Ramirez}, {Sab{\'{\i}}n-Sanjuli{\'a}n},
  {Sim{\'o}n-D{\'{\i}}az}, {Smartt}, {Stroud}, {van Loon}, {Taylor}, \&
  {Vink}}]{dekoter1}
{de Koter}, A., {Sana}, H., {Evans}, C.~J., {et~al.} 2011, Journal of Physics
  Conference Series, 328, 012022

\bibitem[{{de Mink} {et~al.}(2009){de Mink}, {Cantiello}, {Langer}, {Pols},
  {Brott}, \& {Yoon}}]{selma2009b}
{de Mink}, S.~E., {Cantiello}, M., {Langer}, N., {et~al.} 2009, \aap, 497, 243

\bibitem[{{de Mink} {et~al.}(2013){de Mink}, {Langer}, {Izzard}, {Sana}, \& {de
  Koter}}]{selma}
{de Mink}, S.~E., {Langer}, N., {Izzard}, R.~G., {Sana}, H., \& {de Koter}, A.
  2013, \apj, 764, 166

\bibitem[{{de Mink} {et~al.}(2014){de Mink}, {Sana}, {Langer}, {Izzard}, \&
  {Schneider}}]{selma2014a}
{de Mink}, S.~E., {Sana}, H., {Langer}, N., {Izzard}, R.~G., \& {Schneider},
  F.~R.~N. 2014, \apj, 782, 7

\bibitem[{{Duez} \& {Mathis}(2010)}]{2010A&A...517A..58D}
{Duez}, V. \& {Mathis}, S. 2010, \aap, 517, A58

\bibitem[{{Dufton} {et~al.}(2013){Dufton}, {Langer}, {Dunstall}, {Evans},
  {Brott}, {de Mink}, {Howarth}, {Kennedy}, {McEvoy}, {Potter},
  {Ram{\'{\i}}rez-Agudelo}, {Sana}, {Sim{\'o}n-D{\'{\i}}az}, {Taylor}, \&
  {Vink}}]{dufton}
{Dufton}, P.~L., {Langer}, N., {Dunstall}, P.~R., {et~al.} 2013, \aap, 550,
  A109

\bibitem[{{Ekstr{\"o}m} {et~al.}(2012){Ekstr{\"o}m}, {Georgy}, {Eggenberger},
  {Meynet}, {Mowlavi}, {Wyttenbach}, {Granada}, {Decressin}, {Hirschi},
  {Frischknecht}, {Charbonnel}, \& {Maeder}}]{ekstrom2012}
{Ekstr{\"o}m}, S., {Georgy}, C., {Eggenberger}, P., {et~al.} 2012, \aap, 537,
  A146

\bibitem[{{Evans} {et~al.}(2011{\natexlab{a}}){Evans}, {Taylor}, {Sana},
  {H{\'e}nault-Brunet}, {Bagnoli}, {Bastian}, {Bestenlehner}, {Bonanos},
  {Bressert}, {Brott}, {Campbell}, {Cantiello}, {Carraro}, {Clark}, {Costa},
  {Crowther}, {de Koter}, {de Mink}, {Doran}, {Dufton}, {Dunstall}, {Garcia},
  {Gieles}, {Gr{\"a}fener}, {Herrero}, {Howarth}, {Izzard}, {K{\"o}hler},
  {Langer}, {Lennon}, {Ma{\'{\i}}z Apell{\'a}niz}, {Markova}, {Najarro},
  {Puls}, {Ramirez}, {Sab{\'{\i}}n-Sanjuli{\'a}n}, {Sim{\'o}n-D{\'{\i}}az},
  {Smartt}, {Stroud}, {van Loon}, {Vink}, \& {Walborn}}]{2011Msngr.145...33E}
{Evans}, C., {Taylor}, W., {Sana}, H., {et~al.} 2011{\natexlab{a}}, The
  Messenger, 145, 33

\bibitem[{{Evans} {et~al.}(2011{\natexlab{b}}){Evans}, {Taylor},
  {H{\'e}nault-Brunet}, {Sana}, {de Koter}, {Sim{\'o}n-D{\'{\i}}az}, {Carraro},
  {Bagnoli}, {Bastian}, {Bestenlehner}, {Bonanos}, {Bressert}, {Brott},
  {Campbell}, {Cantiello}, {Clark}, {Costa}, {Crowther}, {de Mink}, {Doran},
  {Dufton}, {Dunstall}, {Friedrich}, {Garcia}, {Gieles}, {Gr{\"a}fener},
  {Herrero}, {Howarth}, {Izzard}, {Langer}, {Lennon}, {Ma{\'{\i}}z
  Apell{\'a}niz}, {Markova}, {Najarro}, {Puls}, {Ramirez},
  {Sab{\'{\i}}n-Sanjuli{\'a}n}, {Smartt}, {Stroud}, {van Loon}, {Vink}, \&
  {Walborn}}]{evans}
{Evans}, C.~J., {Taylor}, W.~D., {H{\'e}nault-Brunet}, V., {et~al.}
  2011{\natexlab{b}}, \aap, 530, A108

\bibitem[{{Foreman-Mackey} {et~al.}(2013){Foreman-Mackey}, {Hogg}, {Lang}, \&
  {Goodman}}]{emcee}
{Foreman-Mackey}, D., {Hogg}, D.~W., {Lang}, D., \& {Goodman}, J. 2013, \pasp,
  125, 306

\bibitem[{{Georgy} {et~al.}(2012){Georgy}, {Ekstr{\"o}m}, {Meynet}, {Massey},
  {Levesque}, {Hirschi}, {Eggenberger}, \& {Maeder}}]{2012A&A...542A..29G}
{Georgy}, C., {Ekstr{\"o}m}, S., {Meynet}, G., {et~al.} 2012, \aap, 542, A29

\bibitem[{{Gr{\"a}fener} {et~al.}(2012){Gr{\"a}fener}, {Vink}, {Harries}, \&
  {Langer}}]{grafener2012}
{Gr{\"a}fener}, G., {Vink}, J.~S., {Harries}, T.~J., \& {Langer}, N. 2012,
  \aap, 547, A83

\bibitem[{Gray(1976)}]{Gray}
Gray, D. 1976, The Observation and Analysis of Stellar Photospheres, third
  edition edn. (Cambridge University Press)

\bibitem[{{Groh} {et~al.}(2014){Groh}, {Meynet}, {Ekstr{\"o}m}, \&
  {Georgy}}]{2014A&A...564A..30G}
{Groh}, J.~H., {Meynet}, G., {Ekstr{\"o}m}, S., \& {Georgy}, C. 2014, \aap,
  564, A30

\bibitem[{{Herrero} {et~al.}(1992){Herrero}, {Kudritzki}, {Vilchez}, {Kunze},
  {Butler}, \& {Haser}}]{herrero}
{Herrero}, A., {Kudritzki}, R.~P., {Vilchez}, J.~M., {et~al.} 1992, \aap, 261,
  209

\bibitem[{{Howarth} {et~al.}(1997){Howarth}, {Siebert}, {Hussain}, \&
  {Prinja}}]{howarth}
{Howarth}, I.~D., {Siebert}, K.~W., {Hussain}, G.~A.~J., \& {Prinja}, R.~K.
  1997, \mnras, 284, 265

\bibitem[{{Huang} \& {Gies}(2006)}]{huanggies1}
{Huang}, W. \& {Gies}, D.~R. 2006, \apj, 648, 580

\bibitem[{{Knuth}(2006)}]{knuth2006}
{Knuth}, K.~H. 2006, ArXiv Physics e-prints

\bibitem[{{Krumholz}(2014)}]{2014arXiv1403.3417K}
{Krumholz}, M.~R. 2014, in ``Very Massive Stars in the Local Universe", Ed. J.
  S. Vink, in press (arXiv 1403.3417)

\bibitem[{{Kuiper}(1960)}]{kuiper}
{Kuiper}, N.~H. 1960, Proceedings of the Koninklijke Nederlandse Akademie Van
  Wetenshappen, 63, 38

\bibitem[{{Langer}(2012)}]{langer2012}
{Langer}, N. 2012, \araa, 50, 107

\bibitem[{{Lin} {et~al.}(2011){Lin}, {Krumholz}, \&
  {Kratter}}]{2011MNRAS.416..580L}
{Lin}, M.-K., {Krumholz}, M.~R., \& {Kratter}, K.~M. 2011, \mnras, 416, 580

\bibitem[{{Maeder}(1987)}]{1987A&A...178..159M}
{Maeder}, A. 1987, \aap, 178, 159

\bibitem[{{Maeder} \& {Meynet}(2000)}]{2000ARA&A..38..143M}
{Maeder}, A. \& {Meynet}, G. 2000, \araa, 38, 143

\bibitem[{{Mokiem} {et~al.}(2006){Mokiem}, {de Koter}, {Evans}, {Puls},
  {Smartt}, {Crowther}, {Herrero}, {Langer}, {Lennon}, {Najarro}, {Villamariz},
  \& {Yoon}}]{mokiem}
{Mokiem}, M.~R., {de Koter}, A., {Evans}, C.~J., {et~al.} 2006, \aap, 456, 1131

\bibitem[{{Morel} {et~al.}(2014){Morel}, {Castro}, {Fossati}, {Hubrig},
  {Langer}, {Przybilla}, {Scholler}, {Carroll}, {Ilyin}, {Irrgang}, {Oskinova},
  {Schneider}, {Diaz}, {Briquet}, {Gonzalez}, {Kharchenko}, {Nieva}, {Scholz},
  {de Koter}, {Hamann}, {Herrero}, {Maiz Apellaniz}, {Sana}, {Arlt}, {Barba},
  {Dufton}, {Kholtygin}, {Mathys}, {Piskunov}, {Reisenegger}, {Spruit}, \&
  {Yoon}}]{2014arXiv1408.2100M}
{Morel}, T., {Castro}, N., {Fossati}, L., {et~al.} 2014, in ``New Windows on
  Massive Stars", IAU Symposium 307, Eds. G. Meynet, C. Georgy, J. Groh \& P.
  Stee, Cambridge University Press, in press (arXiv 1408.2100)

\bibitem[{{Penny}(1996)}]{penny}
{Penny}, L.~R. 1996, \apj, 463, 737

\bibitem[{{Penny} \& {Gies}(2009)}]{penny2009}
{Penny}, L.~R. \& {Gies}, D.~R. 2009, \apj, 700, 844

\bibitem[{{Petit} {et~al.}(2013){Petit}, {Owocki}, {Wade}, {Cohen},
  {Sundqvist}, {Gagn{\'e}}, {Ma{\'{\i}}z Apell{\'a}niz}, {Oksala}, {Bohlender},
  {Rivinius}, {Henrichs}, {Alecian}, {Townsend}, {ud-Doula}, \& {MiMeS
  Collaboration}}]{2013MNRAS.429..398P}
{Petit}, V., {Owocki}, S.~P., {Wade}, G.~A., {et~al.} 2013, \mnras, 429, 398

\bibitem[{{Ram{\'{\i}}rez-Agudelo} {et~al.}(2013){Ram{\'{\i}}rez-Agudelo},
  {Sim{\'o}n-D{\'{\i}}az}, {Sana}, {de Koter}, {Sab{\'{\i}}n-Sanjul{\'{\i}}an},
  {de Mink}, {Dufton}, {Gr{\"a}fener}, {Evans}, {Herrero}, {Langer}, {Lennon},
  {Ma{\'{\i}}z Apell{\'a}niz}, {Markova}, {Najarro}, {Puls}, {Taylor}, \&
  {Vink}}]{ramirezagudelo}
{Ram{\'{\i}}rez-Agudelo}, O.~H., {Sim{\'o}n-D{\'{\i}}az}, S., {Sana}, H.,
  {et~al.} 2013, \aap, 560, A29

\bibitem[{{Rosen} {et~al.}(2012){Rosen}, {Krumholz}, \& {Ramirez-Ruiz}}]{rosen}
{Rosen}, A.~L., {Krumholz}, M.~R., \& {Ramirez-Ruiz}, E. 2012, \apj, 748, 97

\bibitem[{{Sana}(2013)}]{2013EAS....64..147S}
{Sana}, H. 2013, in EAS Publications Series, Vol.~64, EAS Publications Series,
  147

\bibitem[{{Sana} {et~al.}(2013){Sana}, {de Koter}, {de Mink}, {Dunstall},
  {Evans}, {H{\'e}nault-Brunet}, {Ma{\'{\i}}z Apell{\'a}niz},
  {Ram{\'{\i}}rez-Agudelo}, {Taylor}, {Walborn}, {Clark}, {Crowther},
  {Herrero}, {Gieles}, {Langer}, {Lennon}, \& {Vink}}]{sana}
{Sana}, H., {de Koter}, A., {de Mink}, S.~E., {et~al.} 2013, \aap, 550, A107

\bibitem[{{Sana} {et~al.}(2011){Sana}, {Le Bouquin}, {De Becker}, {Berger}, {de
  Koter}, \& {M{\'e}rand}}]{sana2011}
{Sana}, H., {Le Bouquin}, J.-B., {De Becker}, M., {et~al.} 2011, \apjl, 740,
  L43

\bibitem[{{Sana} {et~al.}(2014){Sana}, {Le Bouquin}, {Lacour}, {Berger},
  {Duvert}, {Gauchet}, {Norris}, {Olofsson}, {Pickel}, {Zins}, {Absil}, {de
  Koter}, {Kratter}, {Schnurr}, \& {Zinnecker}}]{2014ApJS..215...15S}
{Sana}, H., {Le Bouquin}, J.-B., {Lacour}, S., {et~al.} 2014, \apjs, 215, 15

\bibitem[{{Sim{\'o}n-D{\'{\i}}az} \& {Herrero}(2007)}]{simon}
{Sim{\'o}n-D{\'{\i}}az}, S. \& {Herrero}, A. 2007, \aap, 468, 1063

\bibitem[{{Sim{\'o}n-D{\'{\i}}az} \& {Herrero}(2014)}]{simon2014}
{Sim{\'o}n-D{\'{\i}}az}, S. \& {Herrero}, A. 2014, \aap, 562, A135

\bibitem[{{Slettebak} {et~al.}(1975){Slettebak}, {Collins}, {Parkinson},
  {Boyce}, \& {White}}]{slettebak}
{Slettebak}, A., {Collins}, II, G.~W., {Parkinson}, T.~D., {Boyce}, P.~B., \&
  {White}, N.~M. 1975, \apjs, 29, 137

\bibitem[{{Tan} {et~al.}(2014){Tan}, {Beltran}, {Caselli}, {Fontani}, {Fuente},
  {Krumholz}, {McKee}, \& {Stolte}}]{TBC14}
{Tan}, J.~C., {Beltran}, M.~T., {Caselli}, P., {et~al.} 2014, in Protostars and
  Planets VI, in press (arXiv: 1402.0919)

\bibitem[{{Ud-Doula} {et~al.}(2009){Ud-Doula}, {Owocki}, \&
  {Townsend}}]{2009MNRAS.392.1022U}
{Ud-Doula}, A., {Owocki}, S.~P., \& {Townsend}, R.~H.~D. 2009, \mnras, 392,
  1022

\bibitem[{{Vink} \& {de Koter}(2005)}]{vink2005}
{Vink}, J.~S. \& {de Koter}, A. 2005, \aap, 442, 587

\bibitem[{{Vink} {et~al.}(2001){Vink}, {de Koter}, \&
  {Lamers}}]{2001A&A...369..574V}
{Vink}, J.~S., {de Koter}, A., \& {Lamers}, H.~J.~G.~L.~M. 2001, \aap, 369, 574

\bibitem[{{Vink} {et~al.}(2011){Vink}, {Gr{\"a}fener}, \&
  {Harries}}]{vink2011a}
{Vink}, J.~S., {Gr{\"a}fener}, G., \& {Harries}, T.~J. 2011, \aap, 536, L10

\bibitem[{{Wade} {et~al.}(2014){Wade}, {Grunhut}, {Alecian}, {Neiner},
  {Auri{\`e}re}, {Bohlender}, {David-Uraz}, {Folsom}, {Henrichs}, {Kochukhov},
  {Mathis}, {Owocki}, {Petit}, \& {Petit}}]{2014IAUS..302..265W}
{Wade}, G.~A., {Grunhut}, J., {Alecian}, E., {et~al.} 2014, in ``Magnetic
  fields throughout stellar evolution", IAU Symposium, Eds. P. Petit, M.
  Jardine \& H. Spruit, Cambridge University Press, Vol. 302, 265

\bibitem[{{Wade} {et~al.}(2012){Wade}, {Ma{\'{\i}}z Apell{\'a}niz}, {Martins},
  {Petit}, {Grunhut}, {Walborn}, {Barb{\'a}}, {Gagn{\'e}},
  {Garc{\'{\i}}a-Melendo}, {Jose}, {Moffat}, {Naz{\'e}}, {Neiner}, {Pellerin},
  {Penad{\'e}s Ordaz}, {Shultz}, {Sim{\'o}n-D{\'{\i}}az}, \&
  {Sota}}]{2012MNRAS.425.1278W}
{Wade}, G.~A., {Ma{\'{\i}}z Apell{\'a}niz}, J., {Martins}, F., {et~al.} 2012,
  \mnras, 425, 1278

\bibitem[{{Walborn} {et~al.}(2014){Walborn}, {Sana}, {Sim{\'o}n-D{\'{\i}}az},
  {Ma{\'{\i}}z Apell{\'a}niz}, {Taylor}, {Evans}, {Markova}, {Lennon}, \& {de
  Koter}}]{walborn2014}
{Walborn}, N.~R., {Sana}, H., {Sim{\'o}n-D{\'{\i}}az}, S., {et~al.} 2014, \aap,
  564, A40

\bibitem[{{Weidner} \& {Vink}(2010)}]{weiderandvink2010}
{Weidner}, C. \& {Vink}, J.~S. 2010, \aap, 524, A98

\bibitem[{{Wickramasinghe} {et~al.}(2014){Wickramasinghe}, {Tout}, \&
  {Ferrario}}]{2014MNRAS.437..675W}
{Wickramasinghe}, D.~T., {Tout}, C.~A., \& {Ferrario}, L. 2014, \mnras, 437,
  675

\bibitem[{{Woosley}(1993)}]{1993ApJ...405..273W}
{Woosley}, S.~E. 1993, \apj, 405, 273

\bibitem[{{Woosley}(2011)}]{2011arXiv1105.4193W}
{Woosley}, S.~E. 2011, in ``Gamma-Ray Bursts", Cambridge Astrophysics Series,
  Eds. C. Kouveliotou, R. Wijers \& S. Woosley, Cambridge University Press

\bibitem[{{Woosley} \& {Heger}(2006)}]{woosley}
{Woosley}, S.~E. \& {Heger}, A. 2006, \apj, 637, 914

\bibitem[{{Yoon} \& {Langer}(2005)}]{yoon2005}
{Yoon}, S.-C. \& {Langer}, N. 2005, \aap, 443, 643

\bibitem[{{Yusof} {et~al.}(2013){Yusof}, {Hirschi}, {Meynet}, {Crowther},
  {Ekstr{\"o}m}, {Frischknecht}, {Georgy}, {Abu Kassim}, \&
  {Schnurr}}]{2013MNRAS.433.1114Y}
{Yusof}, N., {Hirschi}, R., {Meynet}, G., {et~al.} 2013, \mnras, 433, 1114

\bibitem[{{Zahn}(1975)}]{Zahn75}
{Zahn}, J.-P. 1975, \aap, 41, 329

\bibitem[{{Zahn}(1977)}]{Zahn77}
{Zahn}, J.-P. 1977, \aap, 57, 383

\bibitem[{{Zinnecker} \& {Yorke}(2007)}]{zinnecker2007}
{Zinnecker}, H. \& {Yorke}, H.~W. 2007, \araa, 45, 481

\end{thebibliography}

\begin{appendix}

\section{A Bayesian derivation of the intrinsic rotation distribution}\label{app: bayes}

\subsection{Analytical form of the \veq\ distribution}
\label{appendix:A}

As described in the main text, we adopt a two-component general analytical form for the probability density function (PDF) of the intrinsic rotational velocity distribution. Following \citet{ramirezagudelo}, the  adopted PDF is composed of a gamma distribution and a normal distribution (see Eq.~\ref{pdf_ve}). The five free parameters in Eq. \ref{pdf_ve} ($\alpha, \beta, \mu, \sigma, I_\gamma$) also allow us to represent PDFs with a variety of shapes, e.g., a velocity peak with varying skewness or peak location, and a high-velocity tail that can vary from negligible to strong. For example, in the deconvolved rotational velocity distribution of the sample of single O-type stars from the VFTS \citep{ramirezagudelo}, the gamma distribution allowed to represent a low-velocity peak and the normal distribution was used to model an additional high-velocity contribution.

\subsection{$\sin{i}$ distribution}
\label{appendix:B}

We use the intrinsic orbital parameters distributions of the VFTS binary sample and the VFTS binary detection probability as a function of orbital period, mass-ratio, eccentricity, sampling and accuracy of the RV measurements obtained in \citet{sana} to compute the intrinsic distribution of the orbital inclinations of the {\it detected} binaries in the VFTS sample. The obtained inclination distribution shows a quasi-absence ($<5$\%) of systems with $i < 20$\degr\ and an over abundance of systems with  $i > 50$\degr\ compared to a distribution computed with random orientation of the binary plane in the 3D space (Fig.~\ref{fig:sini_dist}). While the difference between the random-orientation distribution and the detected-binary distribution are real, the overall effect is small enough that we can show that -- within the quality of our data  -- our results do not depend of which distribution is adopted. For consistency, we nevertheless proceed by adopting the detected-binary distribution for the binary sample and the random-orientation one for the single stars.

\subsection{$\veq \sin{i}$ distribution}
\label{appendix:C}

Given a PDF for the intrinsic rotational velocity distribution, the associated projected rotational velocity distribution will depend on the distribution of orientations of the rotation axis with respect to the line of sight. In what follows, we consider two different PDFs for the $\sin{i}$ distribution. The PDF for a case where the rotation axes are randomly oriented can be written as: \\

\begin{equation}
h(\sin{i}) = \frac{1}{(1/\sin^2{i} - 1)^{1/2}}.
\end{equation}

\begin{figure}
\centering
\includegraphics[width=\columnwidth]{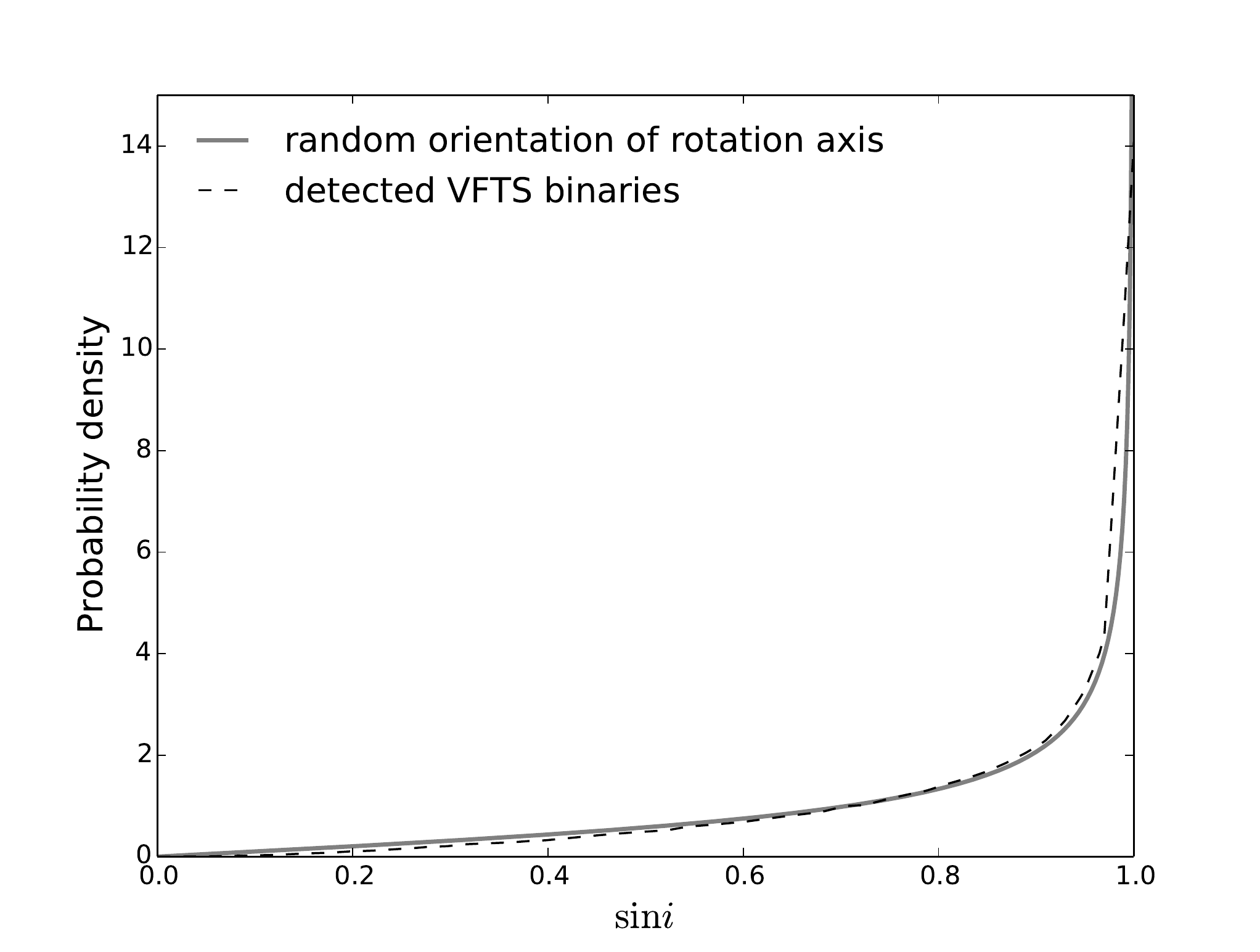}
\caption{$\sin{i}$ distribution for the two different cases considered: a random orientation for the rotation axis, or a rotation axis aligned with the binary axis which is computed for the detected spectroscopic binaries in the VFTS sample.}
\label{fig:sini_dist}
\end{figure}

\noindent{For the sample of primaries of O-type binaries from VFTS, we also consider the PDF of $\sin{i}$ computed for the detected spectroscopic binary systems in the VFTS 
(see Sect.~\ref{appendix:B}). Both $\sin{i}$ distributions considered are actually very similar, as shown in Fig.\ref{fig:sini_dist}}.

The PDF of the projected rotational velocity, $q(\veq \sin{i})$, is obtained by convolving the intrinsic rotational velocity PDF with the $\sin{i}$ distribution in the following way:

\begin{equation}
q(\veq \sin{i}) = \int_{\veq \sin{i}}^{\infty} f(v') \ h\left(\frac{\veq \sin{i}}{v'}\right)\ \frac{1}{v'} \ dv'
\end{equation}

\noindent{where $f$ is as defined in Eq.~\ref{pdf_ve}, and $h$ can be either of the two $\sin{i}$ PDFs discussed above. The likelihood of an individual $\veq \sin{i}$ measurement given the above model and a measurement uncertainty $\sigma_v$ is then obtained by the convolution of the above PDF and a Gaussian distribution (assuming normally distributed errors):}

\begin{equation}
l_{\rm obs}(\veq \sin{i}) = \int_{-\infty}^{\infty} \frac{1}{\sqrt{2 \pi} \ \sigma_v} \exp{\left(\frac{- (\veq \sin{i})^2}{2 \ \sigma_v^2}\right)} \ q(\veq \sin{i} - v') dv',
\label{likelihood}
\end{equation}

\noindent{where $q$ is as defined above.}

\begin{figure}
\centering
\includegraphics[width=\columnwidth]{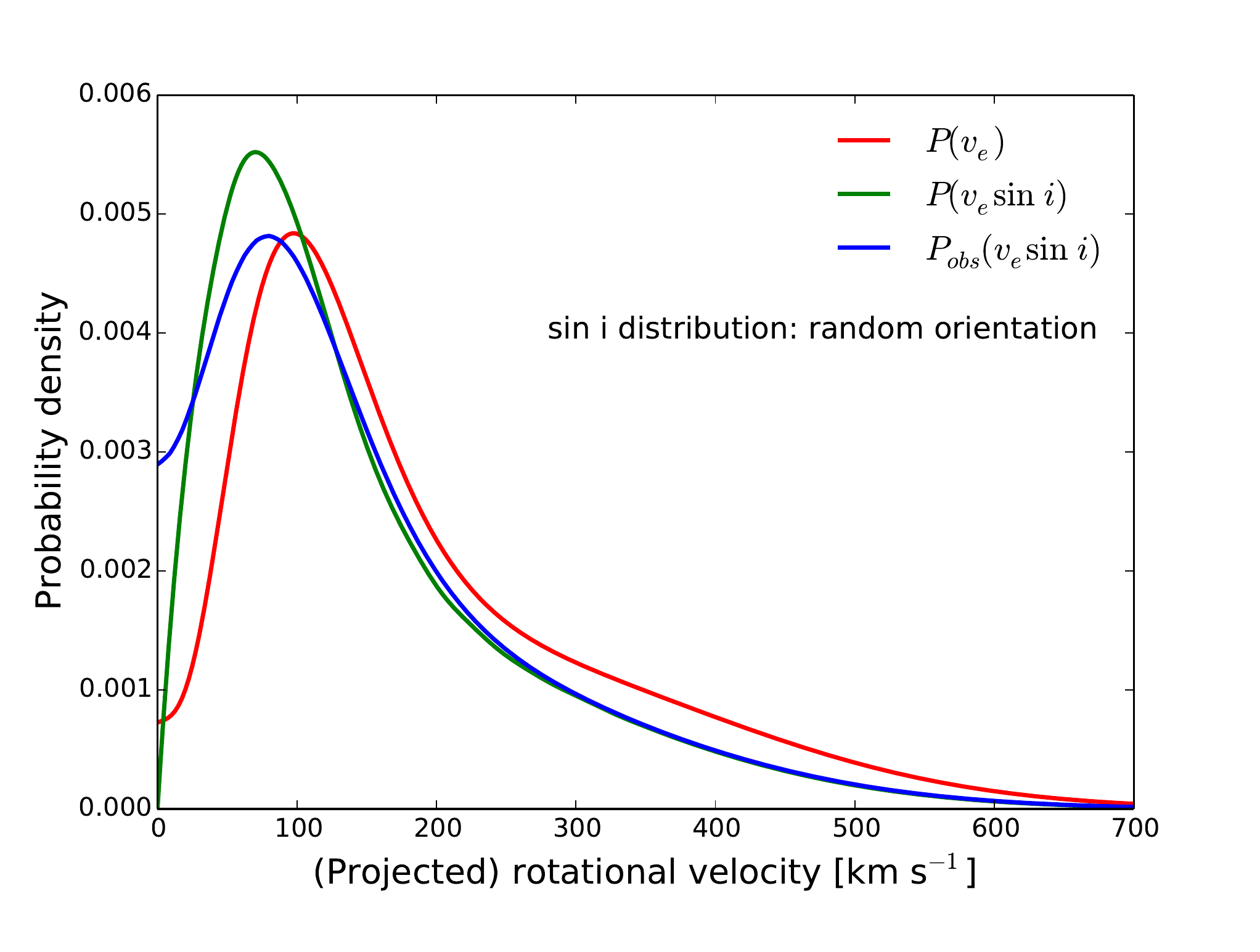}
\includegraphics[width=\columnwidth]{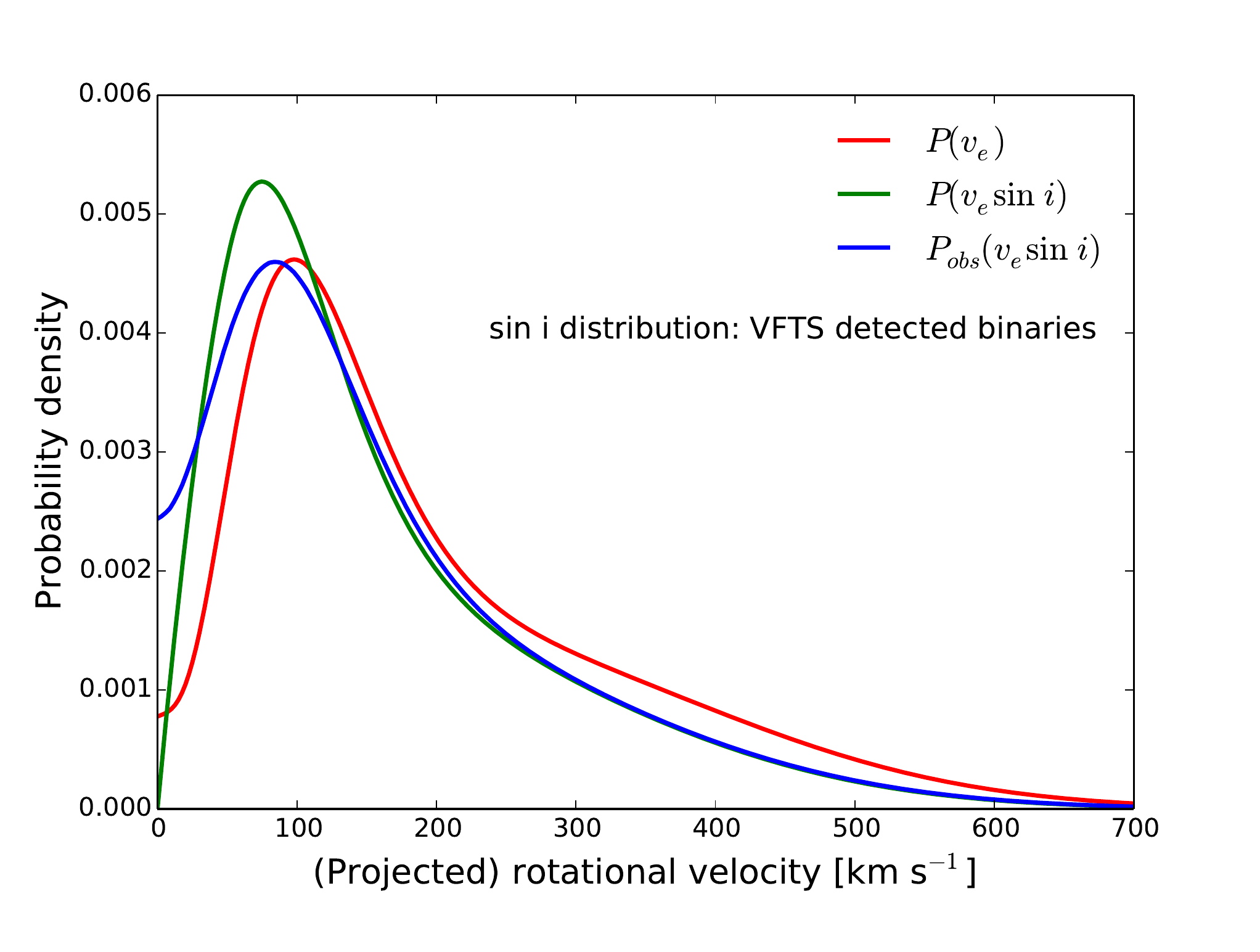}
\caption{PDF of the intrinsic (red), projected (green), and observable (blue) rotational velocity distribution for a model with $\alpha=4.82, \ \beta=1/25, \ \mu=205~\kms, \ \sigma^2= \left(190~\kms \right)^2, \ I_\gamma = 0.43$, and assuming $\sigma_v=20~\kms$. Upper panel: $\sin{i}$ distribution assuming random orientation of the rotation axes. Lower panel: $\sin{i}$ distribution computed for the detected spectroscopic binaries in VFTS is assumed.}
\label{fig:conv}
\end{figure}

To illustrate the effect of inclination and measurement uncertainties on the rotational velocity distribution, we show in Fig.~\ref{fig:conv} the intrinsic, projected, and observable velocity distribution for a chosen set of parameters that were identified by \citet{ramirezagudelo} as providing the best representation of the rotational velocity distribution of the single O-type stars in VFTS ($\alpha=4.82, \ \beta=1/25, \ \mu=205~\kms, \ \sigma= 190~\kms, \ I_\gamma = 0.43$). For this illustration, we assume a measurement uncertainty of $\sigma_v=20 \ \kms$. We show an example adopting each of the $\sin{i}$ distributions discussed above. As expected, in both cases, the projected rotational velocity distribution is shifted to lower velocities due to the effect of inclination, and observational errors slightly broaden the distribution.

\begin{figure*}
\centering
\includegraphics[scale=0.5]{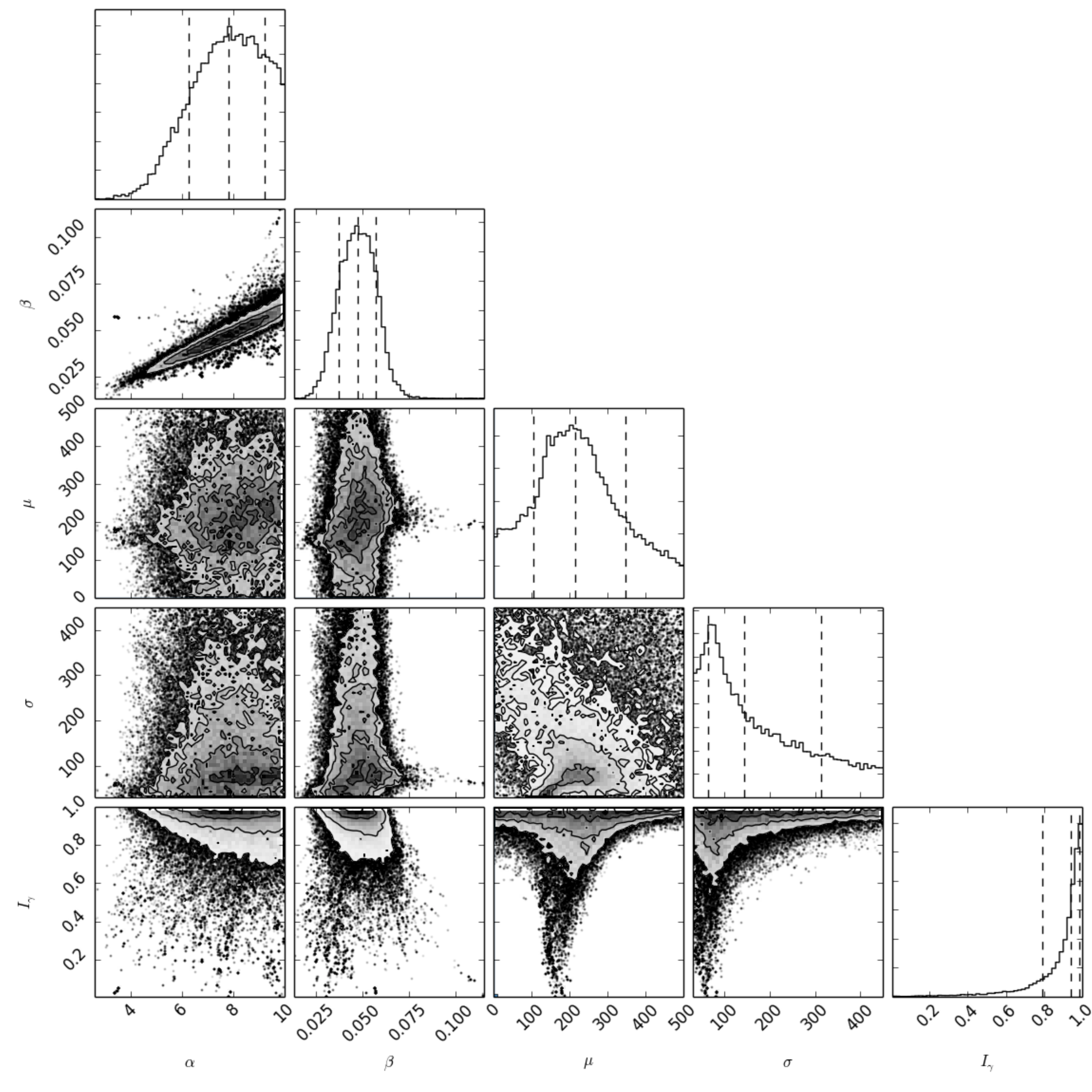}
\caption{One- and two-dimensional projections of the posterior probability distributions of the rotational velocity distribution parameters for the sample of primaries of O-type binaries, assuming the $\sin{i}$ distribution computed for the detected binaries in VFTS.}
\label{fig:post_binary_binarysini}
\end{figure*}

\begin{figure*}
\centering
\includegraphics[scale=0.5]{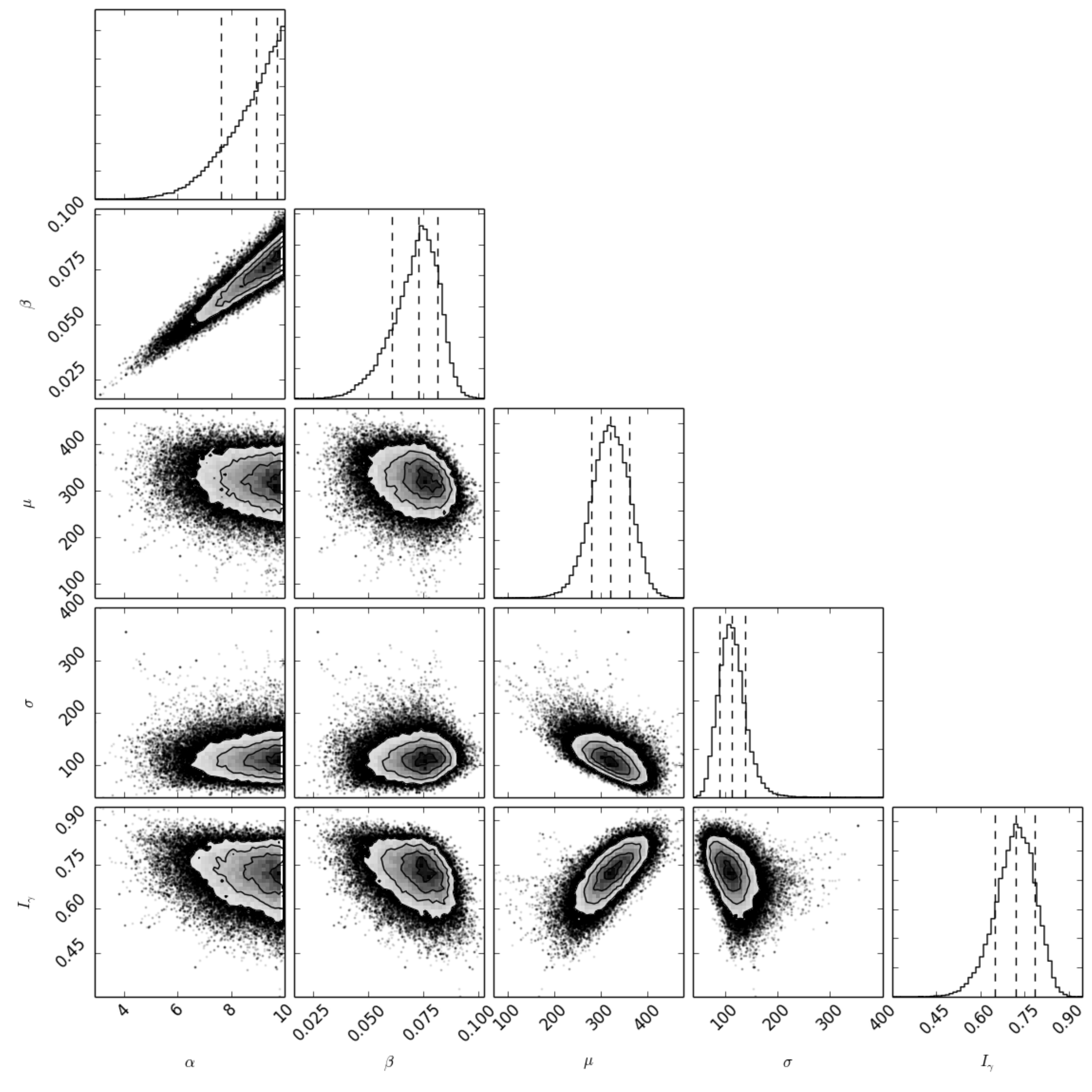}
\caption{One- and two-dimensional projections of the posterior probability distributions of the rotational velocity distribution parameters for the sample of apparently single O-type stars, assuming a random orientation of the rotation axis.}
\label{fig:post_single_randomsini}
\end{figure*}

\subsection{Bayesian analysis and posterior sampling}
\label{appendix:D}

We are interested in the posterior probability density function of the model parameters ($\Theta = \left\{\alpha, \ \beta, \ \mu, \ \sigma, \ I_\gamma \right\}$) given the data (D: a set of $\veq \sin{i}$ measurements with associated uncertainties $\sigma_v$). In a Bayesian framework, the posterior probability density can be written as:

\begin{equation}
p(\Theta | D) = \frac{1}{Z} p(D | \Theta) p(\Theta),
\end{equation}

\noindent{where $p(D | \Theta)$ is the likelihood function and $p(\Theta)$ is the prior distribution. The normalization $Z = p(D)$ is independent of $\Theta$ for a given choice of the form of the generative model, so for the problem that we are interested in here, we can simply sample from $p(\Theta | D)$ without computing $Z$. Our likelihood function is simply the product of individual likelihoods as given by Eq. \ref{likelihood}, i.e.,}

\begin{equation}
p(D | \Theta) = \prod_i  l_{{\rm obs}, i}(\veq \sin{i}_i, \sigma_{v, i}; \alpha, \ \beta, \ \mu, \ \sigma, \ I_\gamma),
\end{equation}

\noindent{where $\veq \sin{i}_i$ corresponds to each individual measurement with uncertainty $\sigma_{v, i}$. We assume uniform priors for all five model parameters over the following ranges: $0.1<\alpha<10, \ 0 < \beta <0.4, \ 0<\mu<500, \ 30 < \sigma < 450, \ 0<  I_\gamma< 1$.}

Finally, to sample efficiently from the posterior distribution and obtain a sampling approximation to the posterior PDF, we use the Python implementation of the affine-invariant ensemble sampler for Markov chain Monte Carlo (MCMC) available through the {\tt emcee} code \citep{emcee}. We use 100 walkers and run the chains for 3500 steps with a burn-in phase of 500 steps. Our final results for the parameters of the rotational probability density functions for the primaries and for the single stars are given in Table~\ref{tab: bayes} and illustrated in Fig.~\ref{fig:vrot_dist_binary_binarysini}.

\end{appendix}

\end{document}